\documentclass{emulateapj}
\usepackage{apjfonts}
\usepackage{lscape}
\input{epsf}

\slugcomment{To appear in the Astronomical Journal. MS \#205339.}

\shorttitle{New Distant Companions to Known Nearby Stars. II.}

\begin{document}

\title{New Distant Companions to Known Nearby Stars. II. Faint
  companions of {\it Hipparcos} stars and the frequency of wide binary
  systems.}

\author{S\'ebastien L\'epine, \& Bethany Bongiorno\altaffilmark{1}}

\affil{Department of Astrophysics, Division of Physical Sciences,
American Museum of Natural History, Central Park West at 79th Street,
New York, NY 10024, USA}

\altaffiltext{1}{National Science Foundation Research Experience for
  Undergraduate intern.}

\begin{abstract}
We perform a search for faint, common proper motion companions of
{\it Hipparcos} stars using the recently published {\em LSPM-north catalog}
of stars with proper motion $\mu>0.15\arcsec$ yr$^{-1}$. Our survey
uncovers a total of 521 systems with angular separations $3 \arcsec <
\Delta \theta < 1500 \arcsec$, with 15 triples and 1 quadruple. Our
new list of wide systems with {\it Hipparcos} primaries includes 130 systems
identified here for the first time, including 44 in which the
secondary star has $V>15.0$. Our census is statistically complete for
secondaries with angular separations $20 \arcsec < \Delta \theta < 300
\arcsec$ and apparent magnitudes $V<19.0$. Overall, we find that at
least $9.5\%$ of nearby ($d<100$ pc) {\it Hipparcos} stars have distant
stellar companions with projected orbital separations $s>1,000$ AU. We
observe that the distribution in orbital separations is consistent
with \"Opik's law $f(s) ds \sim s^{-1} ds$ only up to separation $s
\approx 4,000$ AU, beyond which it follows a more steeply decreasing
power law $f(s) ds \sim s^{-l} ds$ with $l=1.6\pm0.1$. We also find
that the luminosity function of the secondaries is significantly
different from that of the single stars field population, showing a
relative deficiency in low-luminosity ($8<M_V<14$) objects. The
observed trends suggest either a formation mechanism biased against
low-mass companions, or a disruption over time of systems with low
gravitational binding energy.
\end{abstract}




\section{Introduction}

Stellar systems composed of two or more stars (binaries, multiples)
are known to exist in a wide range of configurations. They range from
tight spectroscopic binaries with orbital periods of days to years and
separations under 10 AU \citep{HMUA03}, to marginally bound {\em wide
binaries} with orbital periods $>10^5$ yr and separations up to 20,000
AU (0.1 parsec) or more \citep{CG04}. The frequency of stellar
binaries, and the statistics of their physical properties (periods,
mass ratios, eccentricities) provide constraints on theoretical models
of star formation \citep{B00,KFKM03,DCBH04,GK06}. 

Binaries are very common in the Galaxy, with various estimates of
their frequency hovering around 50\% of all stellar systems.
The distribution in orbital periods $P$ has been estimated over
several orders of magnitude, from days to millions of years, in the
classical papers by \citet{DM91} and \citet{FM92}. The orbital period
distribution of spectroscopic doubles was also re-estimated more
recently by \citet{GML03}. There appears to be a general consensus in
the literature for an orbital period distribution which is uniform in
$\log P$ over several orders of magnitude. The orbital period
distribution is traditionally modeled with a power law $f(P) dP \sim
P^{-1} dP$, sometimes referred to as {\it \"Opik's law} after
\citet{O24}; an equivalent form uses the orbital separation of the
components $f(S) dS \sim S^{-1} dS$. This empirical law breaks down
both at very short orbital periods \citep{GML03}, and at large orbital
separations \citep{CG04}, where the binary frequency is found to drop.

The observed distribution of {\em wide binaries} is of particular
interest, because the low binding energy of these systems makes them
susceptible to disruption. Theoretical models suggest that marginally
bound systems in the Galactic disk will dissolve over timescales
comparable with the age of the Galaxy through repeated, separate
encounters with field stars, giant molecular clouds, or hypothetical
dark matter bodies \citep{RK82,WSW87,MF01}. Since the probability of a
system to become disrupted increases with component separation, it is
expected that an evolved population of wide binaries will exhibit a
break in its distribution at large orbital separations. Hence even a
population whose initial distribution strictly following \"Opik's law
($f(S) dS \sim S^{-1} dS$) will end up with a truncated distribution,
with $f(S)$ breaking down to a steeper power law ($f(S) dS \sim S^{-l}
dS$, with $l>1$) beyond some orbital separation limit. The specific
form of $f(S)$ will be a function of both the age and interaction
history of the binary population under study. Samples of wide binaries
can thus be used as probes of the environments of various populations
of objects. To investigate these statistical effects, it is however
desirable to assemble {\em large samples} of wide binary systems. Much
can also be learned by assembling and comparing samples drawn from a
variety of populations and stellar subtypes \citep{A83,PA04}.

The classical method to find wide binaries is to search for
pairs of stars with small angular separations on the sky. The main
problem is to distinguish between {\em physical pairs}, i.e. genuine
resolved binary stars, and {\em optical pairs}, which are chance
alignments of unrelated stars along the line of sight. While all
physical binaries should ultimately be confirmed through the detection
of their orbital motion, this is generally impractical for wide
binaries with orbital periods of $10^{4}$ years or longer. Wide binaries
are thus generally identified based of probability arguments. Besides
angular separations, other observables are useful is assessing the
status of candidate wide binaries. Parallax measurements are perhaps
the most useful, when they exist; orbital separations are expected to
be $\lesssim1$ pc, so both component should have very nearly identical
parallaxes. Wide binaries are also expected to have orbital velocities
$<1$ km s$^{-1}$, so their radial velocities should be nearly
identical, as well as their proper motion vectors. While radial
velocities are efficient at separating out physical and optical pairs
at any distance range, parallaxes and proper motions are most useful
at small radial distances, where their relative accuracies are higher.

One identifies in the literature two distinct approaches in
identifying wide binaries. One approach attempts to identify distant
wide binaries in deep imaging surveys using statical arguments based
on angular separations alone. A classical example is the two-point
correlation method elaborated by \citet{BS81}. Their analysis of deep
images of North Galactic Pole demonstrated an excess of pairs with
small angular separations, compared to what one would expect from a
random distribution of single stars. Radial velocity measurements have
been used to confirm the binary status of the candidates
\citep{LTBSS84}. This, and other similar surveys
\citep{G88,SG89,GBMY95}, have however yielded only limited samples of
wide binaries.

The second, more successful, approach is based on a search for wide
binaries among catalogs of stars with large proper motions. In such
catalogs, wide binaries are expected to be found as {\em common proper
 motion} (CPM) doubles. This method has been expanded to a
considerable extent by W. J. \citet{L88}. His {\em LDS
  Catalogue}\footnote{CDS-ViZier catalog number I/130} \citep{L87},
contains over 6,200 wide binaries, with proper motions
$\mu>0.1\arcsec$ yr$^{-1}$. Searching for wide binaries among high
proper motion stars has several advantages. Such stars tend to be
relatively nearby ($d\lesssim100$ pc), which means that wide binaries
with orbital separations as small as $a\gtrsim10^2$ AU will generally
be resolved in most imaging surveys, including archived photographic
plate. The proximity of the high proper motion population also makes
it easier to identify {\em intrinsically faint} companions, such as M
dwarfs or brown dwarfs. High proper motion stars are also generally
within range of accurate parallax measurements, which ultimately
provide the most stringent test of the binary star status of the
components.

Unfortunately, a complete catalog of all the high proper motion stars
surveyed for CPM companions was never published by Luyten, leaving no
data on the parent population. In particular, it must be emphasized
that the LDS catalog is not a subsample of the {\em NLTT
  catalog}\footnote{CDS-ViZier catalog number I/98A} \citep{L79}; the
NLTT catalog compiles only the fastest stars identified by Luyten
($\mu>0.18\arcsec$ yr$^{-1}$) and $\approx69\%$ of the LDS pairs have
proper motions below the NLTT limit. Other limitations of both the LDS
and NLTT catalogs include (1) a limited precision in the tabulated
proper motions \citep{GS03} which leaves significant uncertainties on
some of the pairings, and (2) a photographic magnitude system
($m_{pg}$, $m_r$) which does not yield reliable photometric distances
for red dwarfs \citep{W84}. Both the LDS and NLTT catalogs also suffer
from incompleteness, particularly in areas of high stellar densities
such as in Milky Way fields, and also at faint ($V\gtrsim17$)
magnitudes. Nevertheless, the LDS and NLTT catalogs still provide
ample opportunities to conduct additional surveys and analyses of wide
binary systems, and one can argue that they have remained
under-utilized in this respect. A successful use of the Luyten
catalogs is demonstrated by \citet{APH00}, who surveyed a well-defined
sample of $\approx 1,200$ high proper motion objects, identifying and
analyzing 122 CPM companions.

A more systematic search for wide binaries using the Luyten catalogs
was made possible with the release of the revised NLTT catalog (rNLTT)
assembled by \citet{SG03}. The rNLTT is based on the re-identification
of NLTT stars in the USNO-A catalog \citep{Metal98} and in the 2MASS
{\it Second Incremental Release} \citep{Skr97}. This fixed some of the
main limitations of the NLTT by providing significantly more accurate
positions, proper motions, and magnitudes. Additionally, the new
optical-to-infrared colors make it possible to separate out white dwarfs,
disk dwarfs, and halo subdwarfs in a reduced proper motion diagram
\citep{SG02}. A complete analysis of the rNLTT yielded an expanded
list of 1124 CPM doubles \citet{CG04}, including clearly defined
samples of wide binaries from both the local disk and local halo
populations. These samples suggest that the distribution in orbital
separations in both disk and halo wide binaries is consistent with a
power law distribution of the form $f(S) dS \sim S^{-l} dS$ with
$l\simeq1.6$, significantly different from \"Opik's law ($l=1$).

Perhaps the most interesting CPM doubles are those for which at least
one component is a star listed in the {\em Hipparcos} catalog. A
systematic search of both the NLTT and rNLTT by \citet{GC04} \---
hereafter GC04 \--- has yielded 424 such pairs. The identification of a
distant, faint companion to a {\em Hipparcos} star automatically
yields a high-quality parallax for the faint secondary, which can be
critical in mapping the color-magnitude relationship of low-luminosity
stars \citep{Lp05}. But perhaps the most interesting feature for the
investigation of wide binaries is that an accurate parallax also
yields an accurate measurement of the projected {\em physical
 separation} between the two components, which can be more directly
compared with theoretical models.

The new LSPM-north catalog of stars with proper motion
$\mu>0.15\arcsec$ yr$^{-1}$ \citep{LS05} now provides a fresh
opportunity to expand the census of CPM doubles, including systems
with {\em Hipparcos} primaries. The LSPM-north
catalog\footnote{CDS-ViZier catalog number I/298} is based on a
re-analysis of the Digitized Sky Survey (DSS) with the SUPERBLINK
software, which used an image-subtraction algorithm to achieve a high
detection rate for stars with significant proper motions. The catalog
covers the entire hemisphere north of the J2000 celestial equator
(Decl.$>$0). It is is estimated to be $>95\%$ complete over the
entire northern sky down to an apparent magnitude $V=19$.  In fields
at high galactic latitude ($|b|>20$), its completeness is $>99\%$. The
LSPM-north also has a lower proper motion limit than the NLTT
catalog (0.15$\arcsec$ yr$^{-1}$, compared with 0.18$\arcsec$
yr$^{-1}$). Nearly half (30,616) of its 61,977 recorded high
proper motion stars are not listed in the NLTT. Just like the rNLTT,
the LSPM-north provides more accurate positions, proper motions, and
magnitudes, and also gives optical-to-infrared colors.

In paper I, we demonstrated the potential for using the SUPERBLINK
database to find new CPM companions, reporting the discovery of 5 new
companions of known nearby stars \citep{LSR02a}. In this second paper,
we capitalize on the very high completeness of the LSPM-north to
obtain a statistically complete census of faint CPM companions to
{\it Hipparcos} stars. Our methodology for identifying CPM doubles is
described in \S2, our  resulting census of CPM pairs is presented in
\S3 and 4, a statistically complete sample of wide binaries is defined
and  analyzed in \S5, the main results are summarized in the
conclusion (\S6).

\section{Identification of common proper motion doubles}

\subsection{Genuine CPM doubles versus chance alignments}

Common proper motion (CPM) doubles are, by definition, resolved pairs
of stars which have similar proper motions, both in magnitude and
orientation. They typically have angular separations ranging from
several arcseconds to several arcminutes. They are assumed to be
the resolved components of wide binary systems. As such, they are also
expected to orbit around their common center of mass and display
orbital motion. CPM doubles are thus not expected to have {\em
identical} proper motion vectors, since the orbital motion will
result in a small relative proper motion between the components of the
pair. In most cases, however, the orbital period will be large
enough ($\gtrsim10^4$ yr) that the relative proper motion between the
stars will be negligible on timescales of years to decades. Modern
proper motion measurements will thus be comparable to within the
measurement errors. Pairs of high proper motion stars resulting from a
chance alignment, on the other hand, will generally have different
proper motions, although there remains a small probability that a few
will, by chance, also have similar proper motions (although they may
lie at very different distances). The formal identification of a CPM
pair as a wide binary thus requires one to weight in the probability
for a common proper motion pair to be a chance alignment.

Consider two stars at angular positions $\alpha_1,\delta_1$ and
$\alpha_2,\delta_2$, where $\alpha$ refers to the Right Ascension and
$\delta$ to the Declination on the sky. These stars have proper
motions $\mu_{\alpha_1},\mu_{\delta_1}$ and
$\mu_{\alpha_2},\mu_{\delta_2}$. For small angles of separation
between stars 1 and 2, we can approximate the angular separation
$\Delta\theta$ between the stars as:
\begin{equation}
\Delta\theta\simeq((\alpha_1-\alpha_2)^{2}cos{(\delta_1)}^2+(\delta_1-\delta_2)^{2})^{\frac{1}{2}}.
\end{equation}
We can also define $\Delta\mu$ as the magnitude of the difference
between the proper motion vectors:
\begin{equation}
\Delta\mu=((\mu_{\alpha_1}-\mu_{\alpha_2})^{2}+(\mu_{\delta_1}-\mu_{\delta_2})^{2})^{\frac{1}{2}}.
\end{equation}
Each pair can thus be represented in a ($\Delta\theta$,$\Delta\mu$)
phase space. For large numbers of pairs, one can then statistically
determine the probability for a pair to be a chance alignment by
comparing the actual distribution of pairs in the
($\Delta\theta$,$\Delta\mu$) phase space with a model of the
distribution expected from chance alignments.

The expected distribution of chance alignments in
($\Delta\theta$,$\Delta\mu$) phase space is very hard to modelize
analytically for a proper motion catalog such as the LSPM-north. One
complication is that the distribution of high proper motion stars is
not uniform in $\alpha$ and $\delta$ with some regions of the sky
being more densely populated that others \citep{LS05}. The
distribution in proper motions $\mu_{\alpha},\mu_{\delta}$ is also not
uniform. Proper motion vectors have preferred orientations in some
parts of the sky, most notably because of the asymmetric drift of
thick disk and halo stars, as well as from the Solar motion relative
to the local standard of rest, both effects resulting in preferred
orientations of the proper motion vectors in specific parts of the
sky. Also, the density of objects increases as the proper motion
decreases, with very high proper motion stars being rare and low
proper motion stars more common.

It is however possible to obtain an empirical model of the
density of chance alignments in $\Delta\theta$,$\Delta\mu$ phase
space. One can picture the calculation of all the $\Delta\theta$ as a
comparison between the positions of all stars in two sets of objects:
set \#1 with positions $\alpha_1,\delta_1$, and set \#2 with positions
$\alpha_2,\delta_2$. We calculate the angular distance separating 
each individual object in set \#2 from each individual object in set
\#1. Assuming there are $n1$ objects in set \#1 and $n2$ objects in
set \#2, then we have $n1 \times n2$ pairings. What we are interested
in is the special case where the primary star is a {\it Hipparcos}
object. Set \#1 is thus the subset of LSPM-north stars with a
{\it Hipparcos} counterpart (4,839 objects), while subset \#2 comprises the
entire LSPM-north catalog (61,976 objects).

It is possible to modify the dataset and simulate a situation where
no genuine CPM pairs would be found, only chance alignments. Let's
redefine the angular separation $\Delta\theta$ as:
\begin{equation}
[\Delta\theta]_{sim}=( (\alpha_1-[\alpha_2-\Theta])^{2}
  \cos{(\delta_1)}^{2}+(\delta_1-\delta_2)^{2} )^{\frac{1}{2}}.
\end{equation}
This is equivalent to displacing all stars in set \#2 by an angle
$\Theta$ in the direction of Right Ascension, prior to calculating
angular separations. This procedure effectively eliminates all
pairings of genuine CPM doubles (since the second star of the pair is
now moved out the way by some angular distance $\Theta$) while leaving
intact the statistics of the chance alignments. The procedure can be
illustrated with a ballroom analogy. Imagine a crowd of men and women
attending a dance function in a large ballroom. Assume that a few of
the attendees are married couples, while the rest are singles. Let's
say initially all the dancers are scattered around the ballroom at
random, but with all the married couples holding hands. People would
be found in pairs, some because they are married couples, but other
simply by chance. At a specified time, an announcer commands all the
men to walk ten steps to their right. After this, new pairings of
people would have been made, but this time none of them would be
married couples, all would be chance pairings. If we were to calculate
how many pairs there were initially, and subtract out the number of
pairs after the move, then we would have a good estimate of how many
married couples are attending the dance.

The distribution of all pairs in the $\Delta\theta_{sim}$,$\Delta\mu$
thus effectively simulates a distribution of pairs in the LSPM-north
catalog {\em as if there were no CPM doubles in it}. The displacement
angle $\Theta$ should be large enough that all CPM doubles are well
separated, i.e., larger than the typical angular separation between CPM
pairs. However, $\Theta$ should not be too large because we need to
simulate pairing made within areas of the sky that have similar
densities and proper motion systematics. We use values of $\Theta$
between 1 and 5 degrees.

\begin{figure*}
\epsscale{1.0}
\plotone{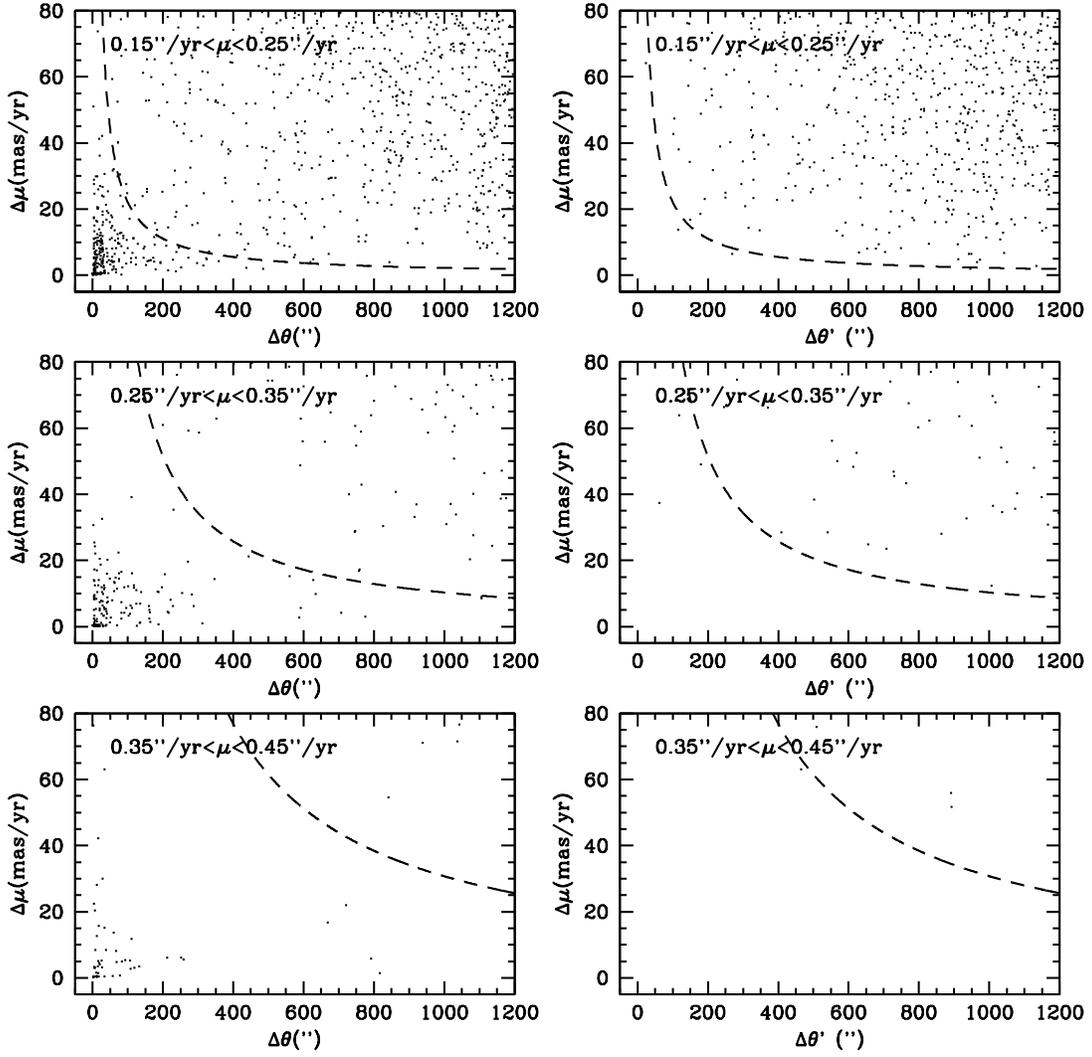}
\caption{Left panels: observed difference in proper motion $\Delta\mu$
  versus angular separation $\Delta\theta$ for pairs of stars in the
  LSPM-north catalog with a {\it Hipparcos primary}. Distributions are
  shown for three different ranges of absolute proper motion (see
  labels); pairs of stars with very high proper motions are much rarer
  because their angular density on the sky is significantly
  lower. Right panels: simulations of $\Delta\mu$, $\Delta\theta$
  distributions for a catalog of high proper motion stars in which no
  wide binaries are present, and pairs occur only from chance
  alignments (see text). The overdensity at small values of
  $\Delta\theta$ and $\Delta\mu$, in the distribution of observed
  pairs, are clearly due to wide binary systems, as it cannot be
  reproduced by chance alignments alone. The dashed lines show the
  approximate location of our adopted limits for the selection of
  candidate wide binary systems.}
\end{figure*}

In Figure 1, we compare the distribution in $\Delta\theta$,$\Delta\mu$
phase space for LSPM-North pairs with distributions of pairs in the
simulated $\Delta\theta_{sim}$,$\Delta\mu$ phase space. The left
panels show the distribution of pairs with {\it Hipparcos} primaries, while
the right panels shows simulations calculated with $\Theta$ = 1
degree. The distributions are split in three different groups,
depending on the mean proper motion $\mu$ of the pairs, 
$0.15\arcsec$yr$^{-1}<\mu<0.25\arcsec$yr$^{-1}$ (top),
$0.25\arcsec$yr$^{-1}<\mu<0.35\arcsec$yr$^{-1}$ (middle), and
$0.35\arcsec$yr$^{-1}<\mu<0.45\arcsec$yr$^{-1}$ (bottom). The panels
to the right display the results of the chance alignment simulations
(i.e. with set \#2 translated by an angle $\Theta$), and for
comparable ranges of $\mu$. 

The most striking difference between the data and the simulations are
the number of pairs found in the lower left corner of each
diagram, corresponding to small angular separations {\em and} small proper
motion differences. Very few pairs are found in the simulations, but
the data shows a large concentration of objects in that range. This
concentration of objects in the data can only be interpreted as the
signature from a population of common proper motion binaries. Pairs of
stars found in that region of the parameter space are thus most likely
to be genuine physical doubles. The density of chance alignments,
however, progressively increases for larger values of $\Delta\theta$
and $\Delta\mu$. There exists a limit beyond which the density of
chance alignment pairs exceeds that of the CPM doubles. This limit can
be used to objectively define a search area for wide binary systems.

\subsection{Objective identification of CPM doubles}

The distribution of high proper motion stars is locally uniform in
$\alpha,\delta$, so the probability of having a chance
alignment, for small values of $\Delta\theta$, is directly
proportional to $\Delta\theta$. On first order, one also expects the
probability of a chance alignment to increase proportionally to
$\Delta\mu$, again for small values of $\Delta\mu$. Thus, for small
values of both $\Delta\theta$ and $\Delta\mu$ the probability of a
chance alignment is expected to be proportional to
$\Delta\mu\Delta\theta$. In the LSPM-North catalog, the
distribution of high proper motion stars empirically follows
N($\mu$)d$\mu\simeq\mu^{-2.8}$d$\mu$, where
$\mu=\sqrt{(\mu_{\alpha})^2+(\mu_{\delta})^2}$. From that, we infer
that the number density of chance alignments $N_{sim}$ in
$\Delta\theta,\Delta\mu$ phase space must follow:
\begin{equation}
N_{sim}(\Delta\theta,\Delta\mu)\simeq\mu^{-3.8}\Delta\theta\Delta\mu
\end{equation}
for small values of $\Delta\theta$ and $\Delta\mu$.

\begin{figure}[t]
\epsscale{1.6}
\plotone{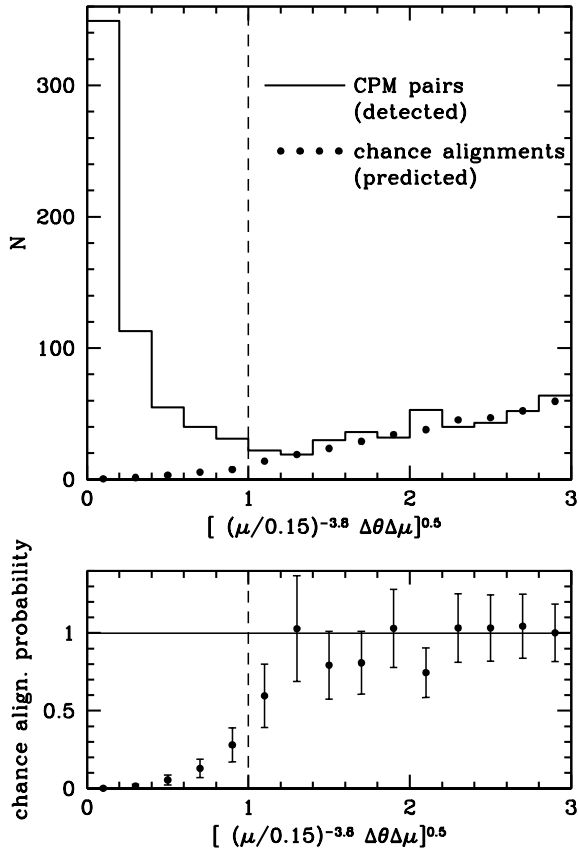}
\caption{Top: comparison between the number density of common proper motion
  pairs in the LSPM-north catalog (histogram) and the predicted number
  density of chance alignments, based on our simulations (filled
  circles). Bottom: ratio between the number of predicted chance
  alignment and the number of pairs found, giving the probability for
  a CPM pair to be a chance alignment. The dashed line marks the
  adopted detection limit for CPM doubles.}
\end{figure}

To determine the range within which a data pair is most likely to be a
CPM double, we define the variable:
\begin{equation}
\Delta X=[(\mu/0.15)^{-3.8}\Delta\theta\Delta\mu]^{0.5},
\end{equation}
which includes a normalization of the proper motions to the lower
limit of the LSPM-north catalog (0.15$\arcsec$ yr$^{-1}$). This new
variable is deliberately set up so that the number density of chance alignment
pairs increases {\em linearly} with $\Delta X$ (hence the square root
in Eq. 5). This helps in modelizing the contribution from chance
alignments, whose density should increase linearly with $\Delta
X$. We compare the number density distribution of pairs
$N_{dat}=N_{dat}(\Delta X)$ in the data with the mean distribution of
simulated, chance alignment pairs $N_{sim}=N_{sim}(\Delta X)$,
averaged from 20 different simulations. The result is displayed in
Figure 2. For small values of $\Delta X$, corresponding to small
angular distances $\Delta\theta$ and/or small differences in proper
motion $\Delta\mu$, one finds that $N_{dat}>>N_{sim}$, which means
that there is a high probability for any pair found in that range to
be a CPM double. We find that $N_{sim}$ increases
linearly with $\Delta X$, as expected, while $N_{dat}$ steadfastly
decreases when $\Delta X$ increases, up to a point where $N_{sim}$
begins to dominate. The bottom panel in Figure 2 estimates the
probability for a pair to be a chance alignment, as a function of
the variable $\Delta X$. We find that beyond $\Delta X=1$ the
probability of a pair to be a chance alignment exceeds 50\%. We thus
use $\Delta X=1$ as the upper limit in our search for CPM doubles.

We select as wide binary candidate any pair for which $\Delta X<1$,
which is equivalent to selecting pairs which satisfy the condition:
\begin{equation}
 \Delta\theta\Delta\mu<(\mu/0.15)^{3.8}
\end{equation}
where $\Delta\theta$ is expressed in seconds of arc, and $\Delta\mu$
is in seconds of arc per year. The location of this boundary is
illustrated in figure 1 for $\mu=0.2\arcsec$yr$^{-1}$ (top),
$\mu=0.3\arcsec$yr$^{-1}$ (middle), and $\mu=0.4\arcsec$yr$^{-1}$
(bottom). Note that the boundary is a function of the mean proper
motion ($\mu$) of the pair, which means that pairs with large proper
motions are selected over a larger range in $\Delta\theta$ and
$\Delta\mu$. 

The boundary rises to high values of $\Delta\mu$ at very
small angular separations, setting an unrealistically high limit for
the detection of wide binaries, whose proper motions should be very
similar in any case. We thus add the following constraint:
\begin{equation}
 \Delta\mu<100mas .
\end{equation}
This large, arbitrary limit ensures that no pairs with very different
proper motions will make the cut, even if they happen to have very
small angular separations. The majority of pairs found in the
LSPM-north are at distances $d>10$pc, and have projected separations
$s>100$AU (see \S 3,4 below); a difference in proper motion
$\Delta\mu>100$mas would imply an orbital velocity $v_{orb}>5$ km
s$^{-1}$ which would be unrealistically large for any system with
$s>100$AU. We also set an upper limit to the range of angular
separations:
\begin{equation}
 \Delta\theta<1,500\arcsec.
\end{equation}
This limits the search to a reasonable range in $\Delta\theta$. Beyond
this limit, the constraints on $\Delta\mu$ from Eq.7 become so
stringent that the allowed range in $\Delta\mu$ falls well below the
measurement errors, thus losing all practicality.

The search area bounded by Eqs.6,7,8 should include the vast
majority of genuine wide binaries, with minimal contamination from
chance alignment pairs. A small level of contamination is still
expected, particularly for pairs selected near the boundaries. Based
on our simulations of chance alignments, we estimate the contamination
in the entire search area to be $\approx10$ pairs, or $<2\%$ of the
total sample.

\section{Wide binaries where both the primary and secondary are
  {\it Hipparcos} stars}

\begin{figure}[t]
\epsscale{2.2}
\plotone{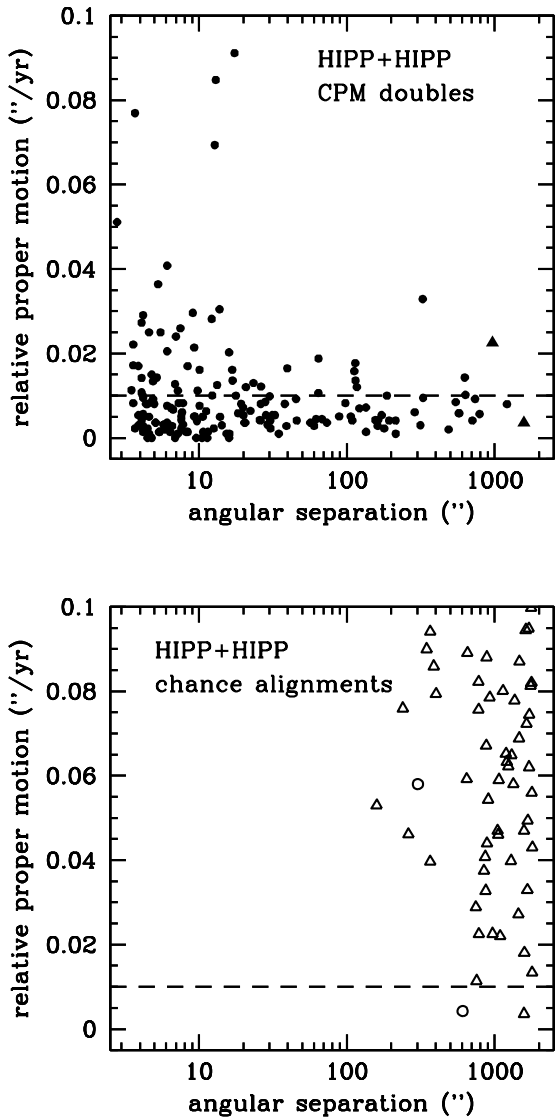}
\caption{Distribution in angular separation $\Delta\theta$ and
  relative proper motion $\Delta\mu$ for pairs composed of two {\it
  Hipparcos} stars. Top panel: pairs identified as wide
  binaries. Bottom panel: pairs identified as probable chance
  alignments. Circles denote the pairs initially selected as common
  proper motion doubles by our search algorithm. Two of these were
  found to have inconsistent {\it Hipparcos} parallaxes, and marked as
  chance alignments (open circles). Triangles denote the pairs
  initially rejected by our search algorithm. Two of the pairs were
  reinstated as candidate wide binaries based on the similarity of
  their parallax measurements (filled triangles).}
\end{figure}

\subsection{Proper motion identification}

Using the selection method discussed above, we searched the
LSPM-north catalog for CPM binaries in which both the primary and
secondary (and tertiary) are {\it Hipparcos} stars. We identified a total of
168 such CPM pairs with angular separations $3\arcsec < \Delta \theta
< 1,200\arcsec$, arranged into 158 doubles and 5 triples. Figure 3
shows the distribution in $\Delta\theta,\Delta\mu$ for the 163 systems
selected as most likely to be wide binaries/multiples (open and filled
circles). Fig.3 also shows the $\Delta\theta,\Delta\mu$ distribution
of pairs that did not fall within our selection limits, and thus
identified as most likely to be chance alignments (open and filled
triangles).

\subsection{Parallax confirmation}

These pairs of {\it Hipparcos} stars make an excellent case study for our
wide binary search algorithm because the {\it Hipparcos} catalog
provides accurate and independent measurements of their parallaxes,
which allow us to verify if the two components indeed physically
associated. Overall, we find that all except 2 of the pair have
the parallax of the primary and secondary within 2.0$\sigma$ of the
measurement errors. For the 2 pairs with mismatched distances, we can
only conclude that they are chance alignments. These pairs are
represented in Fig.3 a open circles (bottom panel).

One of the rejected pairs is formed of LSPM J0722+0849S (HIP 35756)
and LSPM J0722+0854 (HIP 35750), and has an angular separation of
302$\arcsec$. The alleged primary has a {\it Hipparcos} parallax
$\pi_{1}=5.9\pm2.1$mas, while the secondary has
$\pi_{2}=22.4\pm1.6$mas. The two parallax measurements are thus
significantly different. We note that the difference in proper motion
$\Delta\mu$ is also substantial, and it is thus not a surprise that
the two should come out as chance alignments. The pair was initially
selected by our algorithm is mainly because of the relatively large
proper motion of the two stars ($\approx 0.3\arcsec$ yr$^{-1}$), which
makes this alignment a rare event given the relative scarcity of stars
with very large proper motions. This pair was similarly identified as
a chance alignment by \citet{APH00}.

The other rejected pair consists of LSPM J1826+0846 (HIP 90355)
and LSPM J1826+0836 (HIP 90365), with an angular separation of
608$\arcsec$. The first star has a {\it Hipparcos} parallax
$\pi_{1}=37.7\pm1.7$mas, while the second star has a parallax
$\pi_{2}=26.3\pm1.1$mas. Again, the large difference between the two
values indicates that the two are physically unrelated. Here, however,
it comes out more as a surprise because the two stars have proper
motion vectors within only 4 mas yr$^{-1}$ of each other. This
striking alignment of both angular positions and proper motion vectors
serves as a good warning about the possibility of having physically
unrelated stars come up as apparent CPM doubles.

\subsection{Search for additional pairs}

We also examined all pairs rejected in the initial cut, and with
angular separations $\Delta\theta<2,000\arcsec$, to see if
additional CPM binaries could be recovered based on their {\it Hipparcos}
parallaxes. This is one way to check that the algorithm does a good
job at picking out most of the genuine wide binaries. An examination
of the parallaxes of the primaries and secondaries in 55 pairs with
$\Delta\theta<2000\arcsec$ (all rejected on the first pass) reveals
the existence of two pairs that are indeed consistent with being CPM
doubles. More specifically, the parallaxes of the individual
components in each of these pairs are consistent with the two stars
being at the same distance. In each of them, the parallaxes of both
the primary and secondary are better than 10\% accurate, and the
difference between the parallax of the primary and secondary was
smaller than the quoted measurement error. We have added these two
pairs to our list of CPM doubles. In Fig.3, they are plotted as filled
triangles. Note that both pairs have fairly small difference in their
proper motion, which is also consistent with the components forming
physical pairs.

One of the two very wide pairs is composed of the $V=6.0$ star HIP
53791 with the $V=10.8$ star HIP 53756 at a separation
$\Delta\theta=968\arcsec$. The two stars have ${\it Hipparcos}$
parallaxes $\pi=22.0\pm0.8$ and $\pi=22.1\pm2.0$, respectively. The
second pair consists of the $V=6.9$ star HIP 75104 with the $V=9.0$
star HIP 75011. The very large separation between the two components
($\Delta\theta=1582\arcsec$) raises suspicion about their physical
association, but the {\it Hipparcos} parallaxes are a near perfect
match, with $\pi=22.7\pm0.8$ and $\pi=22.3\pm1.2$. At a distance of 44
pc, the projected physical separation between the two components would
be $\sim70,000$ AU.

It is interesting to note that the 2 genuine doubles found among the
55 pairs that were initially rejected suggest that a few percent of
the pairs rejected in our search algorithm are likely to be true
doubles. This emphasizes the limitations of our statistical method in
making a complete identification of wide CPM doubles, especially of
those with very large angular separations."

\subsection{Complete list of wide {\it Hipparcos}+{\it Hipparcos} systems}

Table 1 gives a list of the 158 doubles and 5 triples ultimately 
selected as wide binary systems. The table give the LSPM catalog
number of the primary (column 1), the {\it Hipparcos} catalog number of the
primary (2), its Right Ascension (3), its Declination (4), its proper
motion in the direction or R.A. and in $\arcsec$ yr$^{-1}$ (5), its
proper motion in the direction of Decl. also in $\arcsec$ yr$^{-1}$
(6), its apparent $V$ magnitude (7), its $V-J$ optical-to-infrared
color (8), its parallax in milliarcseconds (9), the estimated parallax
error in milliarcseconds (10), the angular separation between the
primary and secondary in $\arcsec$ (11), all followed by the LSPM catalog
number of the secondary (12), the {\it Hipparcos} catalog number of the
secondary (13), its Right Ascension (14), its Declination (15), its proper
motion in the direction or R.A. and in $\arcsec$ yr$^{-1}$ (16), its
proper motion in the direction of Decl. in $\arcsec$ yr$^{-1}$
(17), its apparent $V$ magnitude (18), its $V-J$ optical-to-infrared
color (19), its parallax in milliarcseconds (20), and finally the estimated
parallax error for the secondary in milliarcseconds (21). In Table 1,
each of the 5 triple systems is represented by two separate (out of the
possible three) pairings. The triples all consist of one short
separation pair with an additional component at larger
separations. The short separation pair is listed in Table 1. The large
separation component is listed in a pair with the brightest component
of the short separation pair. Each triple is thus referenced as one
short period pair and one long period pair.

We find that of the 158 doubles and 5 triples reported here, only 80
of the doubles (and none of the triples) had been previously reported
in GC04. A little more than half the systems in our list are
thus ``new'' CPM systems.

\section{Wide binaries where only the primary is a {\it Hipparcos} star}

\begin{figure}[t]
\epsscale{2.2}
\plotone{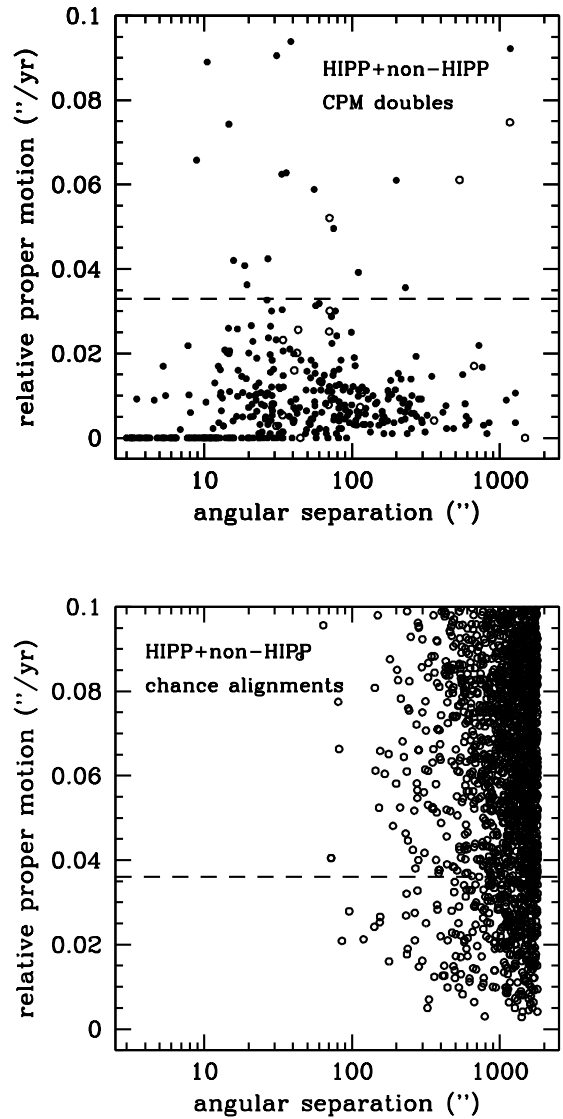}
\caption{Distribution in angular separation $\Delta\theta$ and
  relative proper motion $\Delta\mu$ for pairs for which only the
  primary is a {\it Hipparcos} star. Top panel: pairs identified as
  probable wide binaries by our search algorithm. Open symbols
  denote doubtful pairs, in which the photometric distance of the
  secondary are inconsistent with a physical association. Bottom
  panel: pairs rejected by the algorithm and most likely to be chance
  alignments.}
\end{figure}

\subsection{Identification}

Using again the method described in \S2, we the searched the
LSPM-north catalog for CPM multiples consisting of one {\it Hipparcos}
star with one or more non-{\it Hipparcos} companions. We identified a
total of 347 CPM doubles, 10 triples, and 1 quadruple system. All
systems are listed in Table 2. Figure 4 shows the distribution  in
$\Delta\theta,\Delta\mu$ for all the pairs found with separations
$3\arcsec<\Delta\theta<1,500\arcsec$. Pairs selected as probable CPM
doubles are displayed on the top panel, while the rejected pairs
(probable chance alignments) are shown in the bottom panel.

Note that several pairs with separations $\Delta\theta<100\arcsec$
are found to have relative proper motions $\Delta\mu=0$; this is an
artifact of the LSPM-north catalog. The proper motions listed in the
LSPM-north are based on proper motion measurements of the
stars in the Digitized Sky Survey (DSS), derived with the SUPERBLINK
software \citep{LS05}. The DSS images are scans from the POSS-I and
POSS-II photographic plates. Companions of {\it Hipparcos} stars with
separations $\lesssim20\arcsec$ are generally found within the
extended wings of the bright stars, which are deeply saturated on the
DSS images. In those instances SUPERBLINK identifies the mingled pair
as a single extended, moving object, and calculates a proper motion
for the pair only. The individual components are usually recovered by
visual inspection of either the DSS scans or 2MASS
charts. But even though the stars appear on separate entries in the
LSPM-north, the catalog lists the SUPERBLINK-calculated proper motion
of the {\em pair} for both the primary and secondary. Since their
proper motions are listed as exactly the same, one gets $\Delta\mu=0$
for those pairs. We have verified each of those pairs individually
using our SUPERBLINK-generated finding charts, and confirmed that in
every case the two stars do indeed have very similar proper motions,
so it is very unlikely that any of those pairs could be chance
alignments, and their initial identification as CPM doubles is valid.

\subsection{Photometric distance verification verification}

Because we lack parallax measurements for the non-{\it Hipparcos}
secondaries, it is not possible to verify the binary status of these
pairs by direct comparison of their measured distances, as we did in
\S3. Another test, however, is to check that the color and magnitude
of the secondary is at least consistent with the star being at the
same distance from the Sun as its primary. This is equivalent to
comparing the parallax distance of the primary with the photometric
distance of the secondary. We can plot the secondary star in a
color-magnitude diagram, using the parallax of the {\em primary star}
to calculate its alleged absolute magnitude. If the two stars are
indeed physically associated, then the secondary should fall on the
standard main sequence, or on the white dwarf cooling sequence. If the
secondary stars doesn't fall where it should be on the color magnitude
diagram, then either the secondary is really not at the same distance
from the Sun as the primary, and the pair is a chance alignment, or
the secondary star has an absolute magnitude which doesn't fit with
the standard stellar loci.

\begin{figure*}
\epsscale{1.0}
\plotone{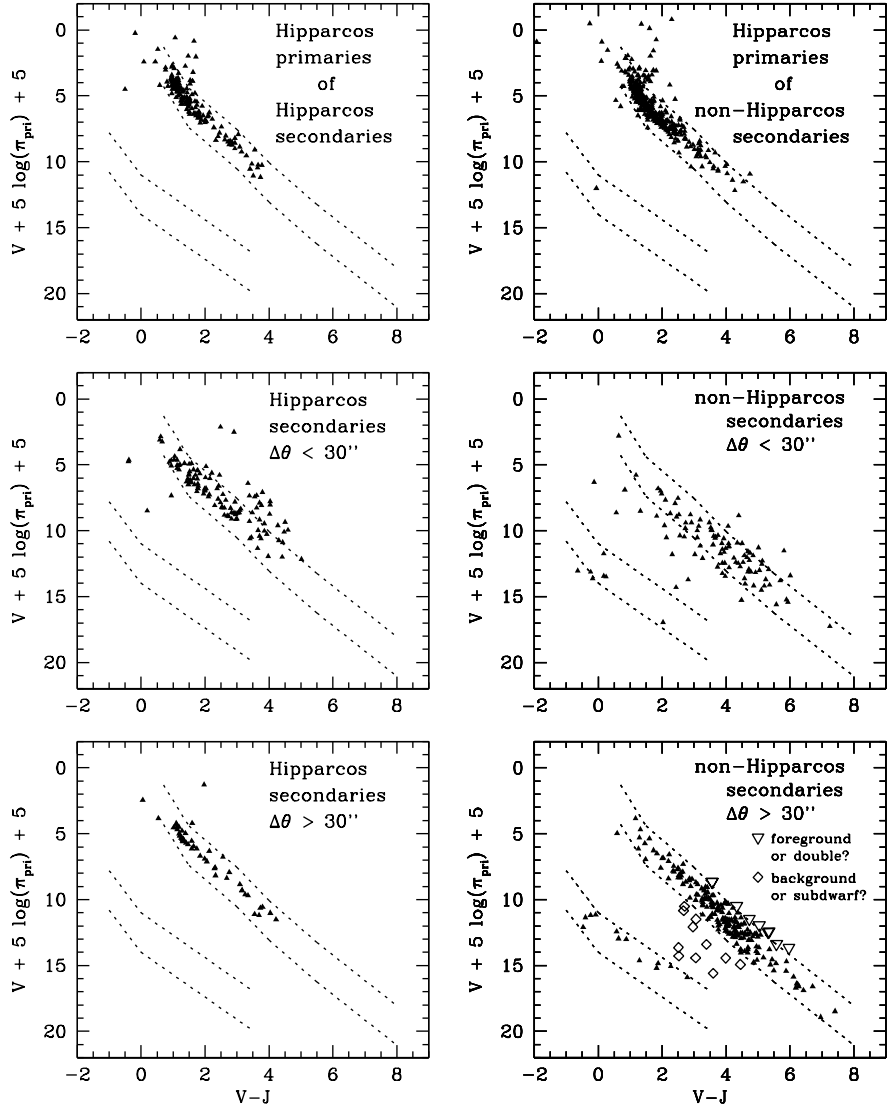}
\caption{Color-magnitude diagrams of the components of the pairs
  identified in our search for wide binaries. Left panels: diagrams
  for the pairs where both components are {\it Hipparcos} stars. Right
  panels: diagrams for the pairs where only the primary is a {\it
  Hipparcos} stars. In all diagrams, the absolute magnitude is
  calculated using the parallax of the primary. Primaries are shown in
  the top row, the middle row shows the secondaries found within
  $30\arcsec$ of their primaries, and the bottom row the secondaries
  found beyond $30\arcsec$ of their primaries. In the tight pairs,
  the secondaries are susceptible to large magnitude errors because of
  the neighboring presence of the very bright primaries. Their location
  on the color-magnitude diagrams are inconclusive. In the wide pairs,
  the magnitudes and colors of the secondaries are more reliable, and
  they are expected to be found within the locus of main sequence
  stars or white dwarfs. Secondaries that fail this test (open
  symbols) may be foreground or background stars, unresolved doubles,
  or subdwarfs.} 
\end{figure*}

Color magnitude diagrams for the primaries and secondaries are
displayed in Figure 5. In these diagrams, the absolute magnitudes of
both the primaries and secondaries are calculated using the parallaxes
of the {\it Hipparcos} primary stars. The diagrams are built in the
(V,V-J) system of visual magnitude and optical-to-infrared color, a
system most useful for displaying the cool end of the main sequence
and perhaps the most appropriate for our sample of secondary stars,
which is dominated by cool red dwarfs. Secondaries found in the
immediate vicinity of bright {\it Hipparcos} stars often suffer from
large measurement errors in their optical V magnitudes, which are
derived from POSS plate measurements. For that reason, we separate the
secondaries into two distinct groups, those with angular separations
$\Delta\theta<30\arcsec$ and those with $\Delta\theta>30\arcsec$. The
color-magnitude test is expected to fail for many of the closer pairs
because of possible large errors in their estimated V
magnitude. Fortunately, pairs with small separations have a very low
probability of being chance alignments, so it is safe to assume that
they are physically associated, in any case.

Figure 5 shows color-magnitude diagrams for the primaries (top), the
close $\Delta\theta<30\arcsec$ secondaries (middle), and the wide
$\Delta\theta>30\arcsec$ secondaries (bottom). The figure also plots
in diagrams in two separate columns, comparing the pairs with {\it
Hipparcos} secondaries from \S2 (left) with the pairs with non-{\it
  Hipparcos} secondaries (right). Overall, we find that the majority
of the secondaries fall close to either the main sequence or the white
dwarf cooling sequence. A pair of dotted lines delimitates a the
region where stars are within 1.5 mag of the standard main sequence
(upper right in the diagram) as determined from a fit of nearby stars
with known trigonometric parallaxes \citep{Lp05}. A second pair of
lines identifies the region within 1.5 mag of the standard white dwarf
sequence based on the distribution of nearby white dwarfs with known
trigonometric parallaxes (L\'epine 2006, in preparation).

Secondaries with $\Delta\theta<30\arcsec$ are found to exhibit
a significantly larger dispersion about the main sequence than the
pairs with $\Delta\theta>30\arcsec$. For the pairs with {\it
  Hipparcos} secondaries, the $\Delta\theta>30\arcsec$ companions have
a mean position of only 0.01 mag above the main sequence with a
$1\sigma$ dispersion of 0.39 mag; the only clear outlier (at
M$_V$=1.27, V-J=1.97) is the star HIP 64522 = HD 115061, a known giant
star. The {\it Hipparcos} secondaries with $\Delta\theta<30\arcsec$,
one the other hand, have many more clear outliers. If we disregard all
the secondaries that are off the main sequence by more than 3
magnitudes, we still find that the companions lie on average 0.42 mag
above the main sequence, with a dispersion of 0.95 mag. This larger
dispersion suggests that the magnitudes and colors of the closer
companions carry large uncertainties, and the 0.42 mag offset even
suggests possible systematic errors. Likewise the magnitudes and
colors of the non-{\it Hipparcos} secondaries are clearly more
scattered around the main sequence for companions with
($\Delta\theta<30\arcsec$). Even if we exclude the 3-magnitude
outliers, we find that the close secondaries are on average 0.11 mag
below the main sequence with a dispersion of 1.33 mag. Companions with
$\Delta\theta>30\arcsec$ have a mean offset of 0.05 mag with a
significantly smaller dispersion of only 0.45 mag. We conclude that
the colors and magnitudes of the short separation pairs
($\Delta\theta<30\arcsec$) are not reliable enough to be useful for a
photometric distance comparison. For the large separation pairs
($\Delta\theta>30\arcsec$), however, the colors and magnitudes are
accurate enough to use the photometric distance of the secondary as a
confirmation that the two components are physically related.

It is clear from Fig.5 that several of the wide secondaries do fail
the test. The lower-right panel shows several secondaries that are
outliers of both the red dwarf and white dwarf sequences. We identify
a total of 20 pairs with mismatched companions; these are listed
separately in Table 3. Nine of these uncertain secondaries are found
lying $\approx2$ magnitudes above the standard main sequence. One
possible explanation is that these secondaries are actually foreground
red dwarfs in chance alignments with background {\it Hipparcos}
stars. Another possibility is that these secondaries are actually
unresolved double stars (in which case these would really be triple
systems). An unresolved pair of stars with components of similar
color and absolute magnitude will appear 0.75 mag brighter than a
single star of the same type. Reducing the absolute magnitudes of
those ``overluminous'' secondaries by 0.75 mag, to account for
unresolved companions, would bring them back within range of the
standard main sequence. The binary status of these stars should
be verified through either high angular resolution imaging
\citep{B04,L06} or radial velocity monitoring \citep{D99}.

The remaining 11 ``uncertain'' companions are found lying between the
main sequence and white dwarf cooling sequence. The simplest
interpretation is that these stars are background red dwarfs in
chance alignments with foreground {\it Hipparcos} stars. They could
also be foreground white dwarfs in chance alignments with background
{\it Hipparcos} stars. Another possibility, however, is that these
secondaries are red subdwarfs. Red subdwarfs are the metal-poor
counterparts of the main sequence red dwarfs; in the color magnitude
diagram, the subdwarf sequence is found to the left (blueward) of the
main sequence \citep{M92,FJ98}. This is because lower metallicities
result in lower atomic and molecular opacities in the optical bands,
shifting the spectral energy distribution to the blue \citep{AH95}. In
the (M$_V$,V-J) color-magnitude diagram, red subdwarfs are expected to
fall between the main sequence and the white dwarf sequence.

\begin{figure}
\epsscale{2.2}
\plotone{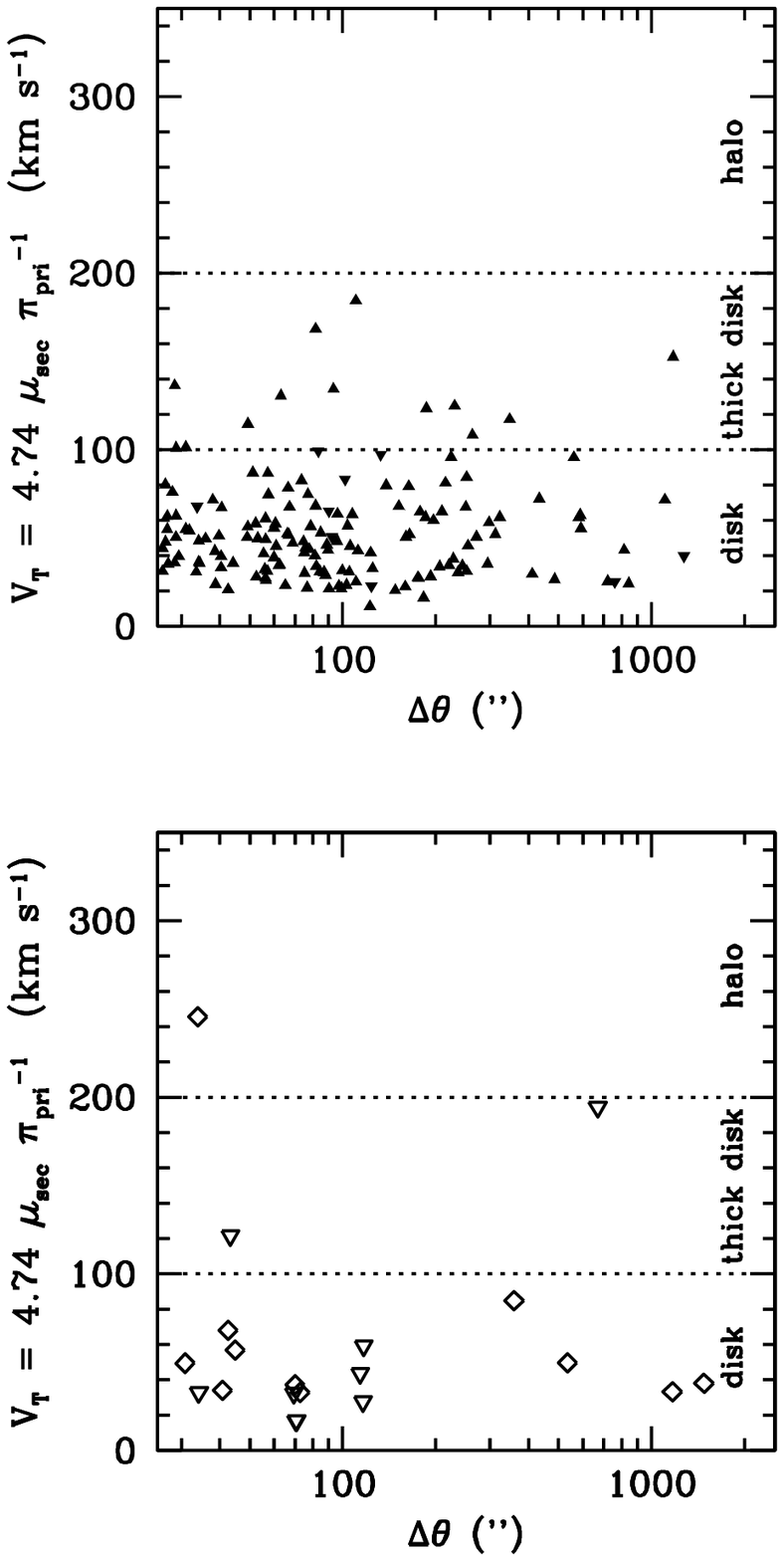}
\caption{Projected, transverse velocities V$_T$ as a function
of angular separation $\Delta \theta$ for pairs in which only the
primary is a {\it Hipparcos} star. Top panel: pairs with matching
photometric distances. Bottom panel: pairs where the secondaries have
inconsistent photometric distances; symbols are the same as in Fig.5.
Of the 13 pairs that could be hosting a subdwarf secondary, only one
has kinematics consistent with the standard subdwarf population (thick
disk/halo). This suggests that the remaining 12 pairs are most likely
to be chance alignments.}
\end{figure}

A simple kinematic test can be applied to check if a star is a good
subdwarf candidate. In the Solar neighborhood, subdwarfs generally
exist as high-velocity stars. This is because local metal-poor stars
are usually members of the Galactic halo or thick disk
\citep{CB00}. In Figure 6, we plot the projected, transverse
velocities V$_T$ of our CPM doubles, against their angular separation
$\Delta\theta$. The top panel shows pairs which have secondaries that
consistently fall on the main sequence when placed at the primary's
distance (see Fig.5). The bottom panel shows the uncertain pairs,
including those for which the secondaries are suspected to be
subdwarfs (using the same symbols as in Fig.5).

We find only one pair with a transverse velocity which clearly
associate it with the halo population. This is the pair whose primary
is the star HIP 35756 (G 89-14) a known spectroscopic double
\citep{LST02}. Although its parallax is fairly uncertain
($\pi=5.9\pm2.1 mas$), G 89-14 has long been known to be a
metal-poor subdwarf, with log(Fe/H)$\approx-1.6$ \citep{SN89}. The
relatively small angular separation to the secondary ($34\arcsec$)
further argues in favor of this pair being a physical double (or
rather triple, since the {\it Hipparcos} primary is itself an
unresolved double). Indeed, this CPM multiple had already been
identified as a physical pair by \citet{APH00}, in their search for
wide binaries among known metal poor stars.

The primaries of the remaining 10 objects, on the other hand, have
kinematics more consistent with the disk population, which makes
the alleged secondaries unlikely to be subdwarfs. For the 4 pairs with
the largest separations $\Delta \theta>300\arcsec$, a good case can be
made that they are most likely to be chance alignments. The situation
is not so clear for the 6 other pairs, which have $30\arcsec <\Delta
\theta< 75\arcsec$: the proximity of the components argues against
simple chance alignments. Either these stars have large errors in
their magnitudes, or they are most likely to be subdwarfs. The
subdwarf status of the alleged CPM companions should be checked more
directly through spectroscopic measurements. Metallicity effects can
be diagnosed in optical spectra from the ratio of the TiO and CaH
bandstrengths \citep{G97,LRS03}. If both the primary and secondary
have spectra consistent with a low metallicity object (i.e. if they
are both spectroscopic subdwarfs) then they are most likely to be
actual physical doubles. If on the other hand, the stars are both
found to be regular dwarfs, then the pair must be a chance
alignment. If one component is found to be a dwarf and the other a
subdwarf, then it would also strongly suggest that the pair is a
chance alignment, since one would expect CPM doubles to have similar
chemical compositions.

\subsection{Overlooked pairs and survey completeness}

Given the angular separation limits of our search algorithm, we do
expect to be overlooking genuine wide binaries with very large angular
separations. The range of our search actually depends on the proper
motion of the pair. For instance, pairs with a mean proper motion
$\mu=0.15\arcsec$ yr$^{-1}$ and proper motion differences
$\Delta\mu=10$ mas yr$^{-1}$ will not make it into the sample if
$\Delta\theta>100\arcsec$. Proper motions in the LSPM-north
catalog have a mean error of 8 mas $^{-1}$, which at first guess
suggests typical proper motion differences $\Delta\mu\approx$11.3 mas
$^{-1}$, for CPM binaries. The actual errors in $\Delta\mu$ are
probably somewhat smaller, because part of the 8 mas $^{-1}$ proper
motion errors in the LSPM-north are due to systematic errors, which
affect both stars in a CPM pair equally, and these systematic errors
subtract out in $\Delta\mu$. In any case, we conservatively adopt 11.3
mas $^{-1}$ as the mean error in $\Delta\mu$, and place a 3-$\sigma$
limit at $\Delta\mu=34$ mas yr$^{-1}$. Since this is larger than our
adopted detection limits at large separations, we conclude that it
is likely that at least a few genuine pairs have been rejected by our
search algorithm.

A look at Fig.4 (bottom panel) shows a handful of pairs with
angular separations $\Delta\theta<300\arcsec$ and relative proper
motions $\Delta\mu<33$ mas yr$^{-1}$ (dashed line) that were not
selected by our algorithm. While these pairs still have a fairly high
probability of being chance alignments, we cannot exclude the
possibility that a significant fraction of them are wide
binaries. Assuming that all the pairs with $\Delta\mu<33$mas yr$^{-1}$
are actually wide binaries, we calculate that these would represent
only $5\%$ of the total pairs found. Hence we are confident that our
census must be at least $95\%$ complete up to
$\Delta\theta=300\arcsec$. Beyond that limit, the completeness of our
survey probably falls as the angular separation increases.

\subsection{Complete list of wide binaries with a {\it Hipparcos}
  primary}

Table 3 gives a list of the 347 doubles, 10 triples, and 1 quadruple
ultimately selected as wide binary systems. The table give the LSPM catalog
number of the primary (column 1), the {\it Hipparcos} catalog number of the
primary (2), its Right Ascension (3), its Declination (4), its proper
motion in the direction or R.A. and in $\arcsec$ yr$^{-1}$ (5), its
proper motion in the direction of Decl. also in $\arcsec$ yr$^{-1}$
(6), its apparent $V$ magnitude (7), its $V-J$ optical-to-infrared
color (8), its parallax in milliarcseconds (9), the estimated parallax
error in milliarcseconds (10), the angular separation between the
primary and secondary in $\arcsec$ (11), all followed by the LSPM catalog
number of the secondary (12), its Right Ascension (13), its
Declination (14), its proper motion in the direction or R.A. and in
$\arcsec$ yr$^{-1}$ (15), its proper motion in the direction of
Decl. in $\arcsec$ yr$^{-1}$ (16), its apparent $V$ magnitude (17),
and its $V-J$ optical-to-infrared color (18). As in Table 1, the
triple systems are represented by two separate (out of the possible
three) pairings. The quadruple system in represented by 3 pairings. A
flag in column 20 identifies the triples and quadruple.

Of the 400 pairings with non-{\it Hipparcos} secondaries, 208 are listed in
GC04; these stars are indicated with a ``GC'' flag in Table 3. Some
134 of these were also listed in Luyten's LDS catalog. We find 26
additional pairs not listed in GC04, but listed in the LDS catalog. The
remaining 130 pairs are new wide binary systems, reported here for the
first time. We note that out of the 156 pairs that are not listed in
GC04, 145 are pairs for which the primary and/or secondary was not
listed in the rNLTT catalog, which explains their absence from the
GC04 list. For the 22 systems whose primaries and secondaries are both
listed in the rNLTT as well as in the LSPM-north, we found one (with
HIP 102040 as the primary) that has a proper motion difference
$\Delta\mu=20.7$ mas yr$^{-1}$ in the rNLTT, which places it just
outside of the selection range used in GC04 ($\Delta\mu<20.0$). The
LSPM-north proper motions yield a $\Delta\mu=11.2$ mas yr$^{-1}$ for that
pair, well inside our own detection range. The other 21 pairs,
however, have $\Delta\Theta$ and $\Delta\mu$ within the selection
limits of GC04, and it remains unclear why they were not included in
their list.

Within the limits of our search area (the northern sky), we found only
4 pairs in the GC04 list that were not picked up by our
search algorithm. These 4 pairs have HIP 45488, HIP 51547, HIP 53008,
and HIP 64386 as their primary stars. Their alleged CPM companions are
relatively distant, with angular separations of 444$\arcsec$,
142$\arcsec$, 178$\arcsec$, and 453$\arcsec$ respectively. We found
differences in proper motion of 21 $mas yr^{-1}$, 24 $mas yr^{-1}$, 16
$mas yr^{-1}$, and 79 $mas yr^{-1}$, respectively, for these pairs;
these fairly large differences, and the large angular separations,
explains why fell outside our selection zone. It is possible that some
of these pairs, and other pairs rejected in our algorithm, might be
genuine physical doubles, particularly those that just barely fell off
the selection limit.

Overall, we find that our own census recovers essentially all the GC04
pairs, and adds a significant number of new objects. The improvement
clearly comes from the use of the more complete LSPM-north
catalog. Given the very high estimated completeness of the LSPM-north,
we are confident that we have now identified virtually {\em all}
existing companions of {\it Hipparcos} stars within a relatively large
range of angular separations, and down to an apparent magnitude
$V=19$.

\section{Analysis of the wide binary census}

\subsection{Distances and orbital separations}

We have determined in \S4.3 that our census is statistically ($>95\%$)
complete up to angular separations $\Delta\theta=300\arcsec$. The
survey also has a limit at small angular separations. The large
apparent magnitude of the {\it Hipparcos} primaries and their deep
saturation on photographic survey plates hampers our detection of
companions with small angular separations. Since the LSPM-north is
based on analyses of Digitized Sky Survey scans of POSS-I and
POSS-II, this is a severe limitation because {\it Hipparcos} stars are
deeply saturated on POSS-I plates. For pairs where only one component
is a {\it Hipparcos} star, we should be able to see companions only
for angular separations $\Delta\theta\ \gtrsim 10\arcsec$. Some
companions in the LSPM-north have been resolved from 2MASS images, but
this is possible only for relatively bright companions, in any case
much brighter than the $V=19$ catalog limit.

Pairs for which both components are {\it Hipparcos} stars are not
limited in the same way, because the LSPM-north uses the {\it Hipparcos}
catalog itself as input for the brighter stars. All pairs with
$\Delta\theta\ \gtrsim 3\arcsec$ should be resolved in the {\it
  Hipparcos} catalog. We can thus use the statistically complete
distribution in angular separations of the {\it Hipparcos}+{\it
  Hipparcos} pairs to estimate the incompleteness of our census for
pairs with faint secondaries.

\begin{figure}
\epsscale{2.2}
\plotone{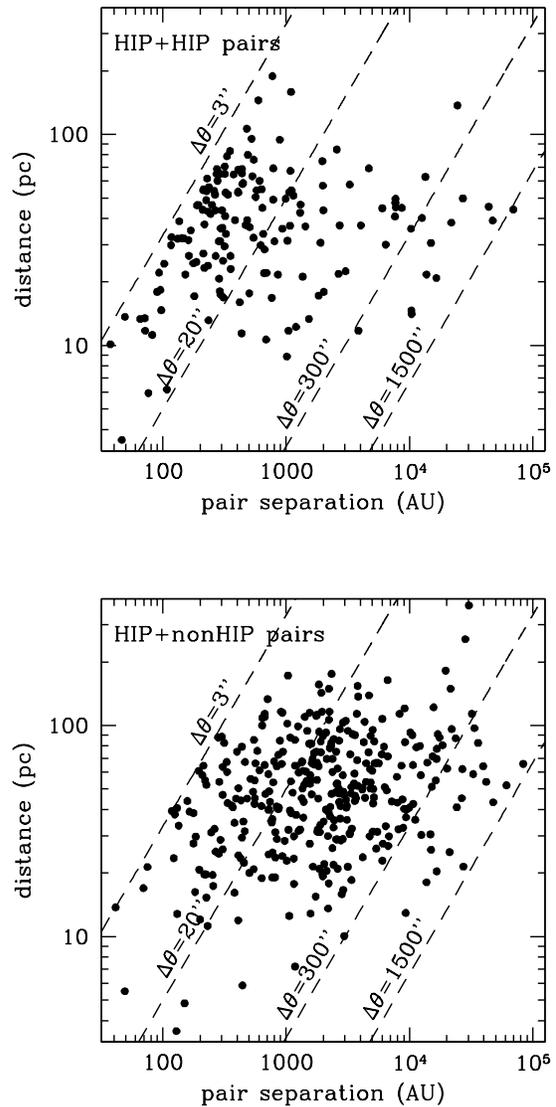}
\caption{Calculated projected physical separations of the wide
  binaries identified in our survey (in Astronomical Units). The
  census is much more complete at small angular separations for
  pairs of {\it Hipparcos} stars (top). For non-{\it Hipparcos}
  companions, the census  appears to be complete only beyond
  $\Delta\theta>20\arcsec$. Pairs are however detected up to
  $\Delta\theta\approx1,500\arcsec$, which gives an excellent coverage
  of the 1,000-10,000 AU separation
  range.}
\end{figure}

Figure 7 plots the projected orbital separations
$s=\pi^{-1}\Delta\theta$ of the pairs as a function of their distance
$d=\pi^{-1}$ (estimated from the {\it Hipparcos} parallaxes
$\pi$). Note that our systems tend to be relatively nearby, with
most of the pairs lying at distances 20-100 parsec. The fact that
fewer pairs are detected at larger distances is the result of proper
motion selection effects. Because of the lower limit $\mu>0.15\arcsec$
yr$^{-1}$ of the LSPM-north, stars with low transverse velocities are
increasingly left out of the sample at larger distances
\citep{Lp05}. The number of detected systems hence reaches a maximum
around $d\approx40$ pc.

What is also apparent in Fig.7 is the dominance of pairs
$\Delta\theta<20\arcsec$ in the HIP+HIP systems. In contrast, pairs
for which only the primary is a {\it Hipparcos} star, are mostly found
with $\Delta\theta>20\arcsec$. Our census of systems with faint
secondaries is thus clearly undersampled in pairs with
$\Delta\theta<20\arcsec$. This is largely consistent with our
prediction that secondaries with small angular separations would be
lost in the glare of their bright primaries. We conclude that our
census is thus statistically complete only over the range
$20\arcsec<\Delta\theta<300\arcsec$.

While our census spans only 1.3 orders of magnitude in $\Delta\theta$,
we are actually sensitive to a larger range of projected orbital
separations $s$ because our sample covers systems over a relatively
large range of distances. Since our pairs are found in a range of
distances 15 pc$<d<$100 pc, our effective sampling range for wide
companions is between 300 AU and 30,000 AU. It is worth noting that
this extends beyond the canonical upper limit $s<0.1$ pc (= 20,000 AU)
for wide binary systems \citep{CRC90}. Our sample actually suggests
the existence of wide binaries with $s>0.1$ pc. Our widest binary
system is the pair composed of HIP 64157 and the faint high proper
motion star LSPM J1308+3332 (= NLTT 33016), which have an angular
separation $\Delta\theta=1273.7\arcsec$ and are located at a distance
$d\approx65$ pc, which yields an orbital separation
$s\approx83,000$AU. Note that the pair has a relatively large
transverse velocity $V_{tran}\approx70$ km s$^{-1}$, which rules out
the alternate interpretation that the two star are co-moving members
of a young association.

\subsection{Absolute magnitudes of the secondaries}

\begin{figure}
\epsscale{2.2}
\plotone{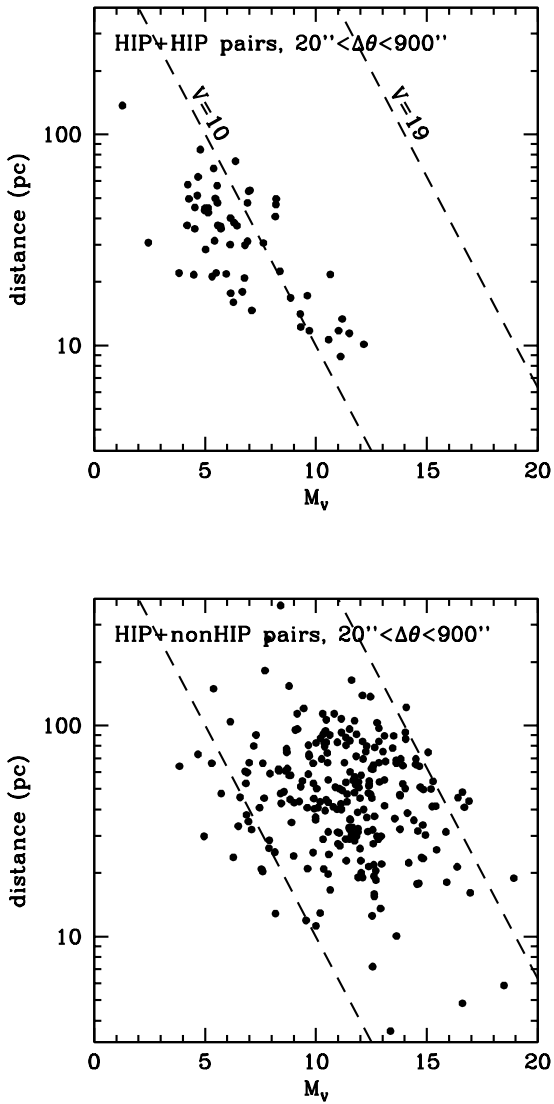}
\caption{Absolute magnitude of the secondaries plotted against the
  distance of the common proper motion systems. Few of the {\it Hipparcos}
  secondaries are found with apparent magnitude $V>10$ due to the
  increased incompleteness in the {\it Hipparcos} catalog (top panel). Many
  more secondaries are found with the LSPM-north (bottom), which is
  statistically complete down to $V=19$. One can see the turnover in
  the luminosity function beyond $M_V=14$, which is similar to the
  turnover in the luminosity function of field single stars.
}
\end{figure}

Figure 8 shows the distribution in absolute magnitudes $M_V$ and
distances of the CPM secondaries, for pairs with separations
$20\arcsec<\Delta\theta<900\arcsec$. In this range of angular
separations, our census should be nearly complete down to the limiting
magnitude of the LSPM-north catalog i.e. $V=19$. We show distinct
plots for {\it   Hipparcos} secondaries (top) and non-{\it Hipparcos}
secondaries (bottom). The secondaries are clearly dominated by stars
with $10<M_V<15$, in other words by M dwarfs. Since most of the
{\it Hipparcos} primaries are G dwarfs, with $5<M_V<8$, it is clear
that there is no general tendency for the secondaries to be of similar
masses as their primaries. The secondaries, in fact, look like they
could have been drawn from the field population of single stars (see
however \S5.3 below).

The magnitude limit of the LSPM-north catalog yields a limiting
absolute magnitude which is dependent on the distance of the
pair. The majority of our pairs are relatively close; if we restrict
our sample to objects within $d<100$ pc, then our census of faint
secondaries should be complete down to $M_V=14$. This happens to be
just beyond the peak in the luminosity function of low-mass red dwarfs
\citep{RGH02}. Assuming the luminosity function of the secondaries to
be roughly similar to that of field stars, then our census should be
missing only a relatively small fraction of the Hydrogen-burning
secondaries. In any case, our detection of several very nearby
($d<20$ pc) secondaries with $M_V>15$ suggests that several very-low
mass companions remain to be discovered at larger distances. To be
detected, these will require catalogs of high proper motion stars with
significantly fainter magnitude limits.

\subsection{Luminosity function of the secondaries}

\begin{figure}
\epsscale{2.5}
\plotone{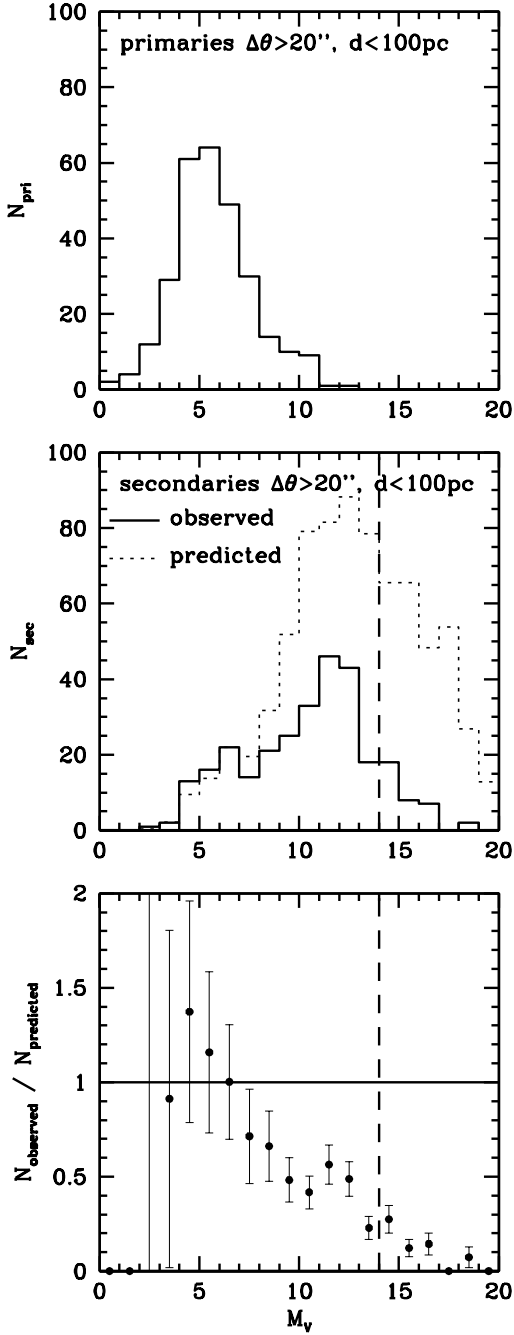}
\caption{Distribution in absolute magnitude $M_V$ for the primaries
  (top) and secondaries (center) in our sample of wide binaries. The
  observed luminosity function of the secondaries is found to be
  significantly deficient in faint stars ($M_V>9$) compared to the
  distribution predicted if both components are draw from the field
  luminosity function of \citep{RGH02}. The observed distribution
  contains only $\approx60\%$ of the predicted number of faint
  secondaries. The dashed line shows the completeness limit of our
  sample ($M_V<14$).
}
\end{figure}

A close examination of the distribution in the absolute magnitude of
the CPM secondaries shows a deficit of low-luminosity objects
($M_V>9$) compared with the field luminosity function. Figure 9 shows
the $M_V$ distribution for the primaries and secondaries in our
sample. We use a Monte-Carlo simulation to predict the magnitude
distribution of the secondaries, using the assumption that both
components are drawn independently from the field luminosity
function.  We adopt the luminosity function of main
sequence stars in the Solar Neighborhood estimated by \citep{RGH02}; 
the additional contribution of the white dwarfs can be neglected,
since they account for less than 5\% of the secondaries. As shown in
Fig.9, the distribution observed for the secondaries is significantly
different from the predicted one. If we adjust the simulations to
match the numbers of bright ($M_V<8$) secondaries, we find only
$\approx60\%$ of objects in the $9<M_V<14$ range predicted by the
model. Since our census is demonstrably complete at $d<100$ pc down to
$M_V=14$, the difference is significant, and rules out the idea that
the luminosity function of the secondaries is comparable to the field
luminosity function.

Since the absolute magnitude of main sequence stars is correlated with
mass, the dearth in low-luminosity objects indicates that low-mass
stars are less likely to be found as CPM secondaries than they are in
the field population. One simple interpretation is that the formation
mechanisms for wide binaries are biased against secondaries of very
low mass or, equivalently, that they are biased in favor of high-mass
secondaries. Another possible interpretation is that wide binaries are
formed initially with both components drawn from the same mass
function, but that gravitational interactions with the environment
disrupt a number of systems over time, in events that favor the
disruption of systems with low-mass secondaries.

In fact, if we assume that both components in wide binaries initially
follow the field luminosity function, then the ratio between the
observed (or ``current'') and predicted (``initial'') luminosity
function (Fig.9) shows a trend that suggest that the lower the mass of
the secondary, the higher the probability for the wide system to have
been disrupted over time. Should this interpretation be correct, then
one should expect to find very few faint companions of {\it Hipparcos}
stars with absolute magnitudes $M_V>14$. This would imply that our
sample is already nearly complete for wide Hydrogen-burning
companions, and that very few fainter companions should be found.
 
%

\subsection{Observed frequency of wide binary systems}

From our statistically complete sample of CPM pairs with {\it
Hipparcos} primaries, we can now estimate the multiplicity fraction
of wide binary systems in the Solar Neighborhood. We have determined
that our northern-sky census was statistically complete for systems
within 100 pc and proper motion $\mu>0.15\arcsec$ yr$^{-1}$, and for
CPM companions with luminosities $M_V<14$. We estimate the
multiplicity fraction by comparing the number of wide binaries found
in our survey to the total number of {\it Hipparcos} stars in the same
range of position, distance, and proper motion. A search of the {\it
  Hipparcos} catalog finds a total of 3,041 stars north of the
celestial equator with proper motion $\mu>0.15\arcsec$, parallaxes
$\pi>10$ mas, and a parallax accuracy of better than 10\%.

We calculate the number density of wide binaries as a
function of the logarithm of the {\em projected}, orbital separation
between the pairs $\log{s}$, with $s$ expressed in AU. Note that $s=S
\sin(i)$, where $S$ is the actual orbital separation between the
components and $i$ the orbital inclination. While $S$ is the
physically meaningful quantity, we only have $s$ as the observable. We
separate our sample into bins of 0.2 in $\log{s}$. We calculate the
multiplicity fraction for each bin by taking the ratio of wide
binaries found in that bin to the total number of {\it Hipparcos}
stars which satisfy the distance detection limits for that bin, based
on a range of angular separations $20\arcsec<\Delta
\theta<900\arcsec$. For instance, the $\log{s}=3.1$ bin can only
comprise systems with distances between 1.4pc and 62.9pc (see Fig 7).
The resulting wide binary frequency distribution is shown in Figure
10. Our results indicate that at least $9.1\pm1.6\%$ of all stellar
systems in the Solar Neighborhood consist of wide binaries/multiples
with separations $s>1,000$ AU. More specifically, this conclusion is
valid for systems in which the primary is a typical, nearby {\it
  Hipparcos} star, i.e. the primary is a G dwarf with absolute
magnitude $4\lesssim M_V \lesssim7$.

\begin{figure}
\epsscale{1.0}
\plotone{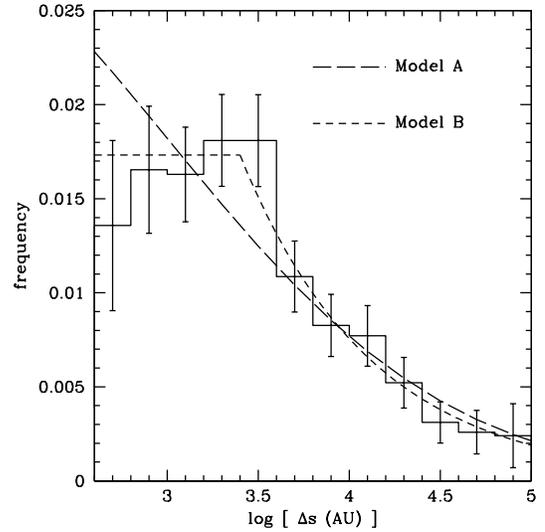}
\caption{The frequency of wide binary systems in the Solar
  Neighborhood, as estimated from our sample of {\it Hipparcos} stars with
  distant common proper motion companions. The frequency is calculated
  for each bin, which samples the projected orbital separation $s$ in
  steps of 0.2. Two possible models are shown: model A is the tail of
  an gaussian, model B follows \"Opik's law ($f(s)\sim \Delta s^{-1}$
  ds) up to 2,500 AU, then breaks down to a power law $f(s)\sim
  \Delta s^{-1.6}$.}
\end{figure}

We try to fit the observed distribution of orbital separation with two
different models previously used in the literature. First we try a
model of the form used by \citet{DM91}, which modelizes the
distribution over the logarithm of orbital periods $\log{P}$ with a
Gaussian:
\begin{equation}
f_P(\log{P}) = F_{mult} \pi^{1/2} \sigma_P \ \exp{-\frac{(\log{P}-P_0)^2}{2
    {\sigma_P}^2}} ,
\end{equation}
with $P_0=4.8$, $\sigma_P=2.3$, $\pi=3.141592...$, and where
$F_{mult}$ is the total multiplicity fraction, i.e.
$F_{mult}=\int_{-\infty}^{\infty} f_P(P) dP$. For the systems
considered in this paper, in which the primary stars are mostly
solar-type objects with masses $M_* \sim 1 M_{\sun}$, we calculate
the following statistical relationship between the period $P$ and
projected separation $s$:
\begin{equation}
2 \log{P} \simeq 3 \log{s} + 5 ,
\end{equation}
where $P$ is expressed in years and $s$ in AU. We then have:
\begin{equation}
f_s(\log{s}) = F_{mult} \pi^{1/2} \sigma_P \ \exp{-\frac{(1.5 \log{s} + 2.5 -
    P_0)^2}{2 {\sigma_P}^2}} .
\end{equation}
Our attempt to fit this form to the observed distribution of
projected separations is shown in Fig.10 (Model A). Our best fit does
not reproduce well the form of the distribution, and predicts more
objects at small separations than are actually observed. In fact, out
best fit model suggests a total multiplicity fraction $F_{mult}=0.75$,
which is much larger than the total binary fraction estimated by
\citet{DM91}. We take this as evidence that this particular form is
inadequate in the range of orbital separations under consideration.

Our second attempt to modelize the distribution of orbital separations
uses a power-law of the form:
\begin{equation}
f_S(S) \sim S^{-l} .
\end{equation}
Given that the mean projected separation in a sample of systems with
random orbital plane orientations is $s\simeq0.78 S$, then
statistically the distribution in projected separations $s$ also
follows:
\begin{equation}
f_s(s) \sim s^{-l} .
\end{equation}
If we plot the distribution along $\log{s}$ we have:
\begin{equation}
f_s(\log{s}) \sim 10^{(1-l) \log{s}}.
\end{equation}
The specific form of the equation for which $l=1$ is the well-known
{\em \"Opik's law}, and yields a distribution which is uniform in
$\log{s}$.

We find that our data can be consistently modeled with two distinct
power-law distributions over two separate ranges of orbital
separations (Fig.10, Model B). At small projected separations, our
data are consistent with {\em \"Opik's law}, but only up to a
separation $s\sim3,000$ AU. Beyond that limit, the distribution shows
a sudden break, and can best be fitted with a $l=1.6$ power index. The
latter is consistent with the trends observed by \citet{CG04} for both
their disk and halo samples. The flattening of the distribution below
$s\sim3,000$ AU, which is also hinted at in the CG04 data, is not much
of a surprise since the $l=1.6$ power law cannot possibly extend to
much smaller orbital separations or the multiplicity fraction might
take unphysical values. An extrapolation to $S=1.4$ ($s=25$AU), for
instance, would yield a total binary fraction larger than 1.

Perhaps the more significant result is not the flattening of the
$l=1.6$ power law distribution toward small angular separations, but
rather the observed break in {\it \"Opik's Law} at large angular
separations. If one believes that star formation processes initially
yield a distribution of wide binary systems consistent with {\it
 \"Opik's Law}, then our data reveals the range of maximum extension
of the law. {\it \"Opik's Law} then probably breaks down because of
the disruptive effects of the environment in which the stars are
born. Disruption would most likely occur through interactions with
neighboring stellar systems, perhaps in the very same cluster in which
these systems are born. A much larger census of wide binary systems,
which would yield a more detailed distribution of orbital separations,
might shed more light on the processes at work in hampering the
formation of extremely wide pairs.

\section{Conclusions}

We have used the LSPM-north catalog of stars with proper motions
$\mu>0.15\arcsec$ yr$^{-1}$ to conduct a search for common proper
motion (CPM) companions of {\em Hipparcos} stars. Because of the large
coverage (20,000 square degrees) and high completeness of the
LSPM-north, we have obtained a large, {\em statistically complete}
census of CPM companions over the range of angular separations
$20\arcsec<\Delta \theta<300\arcsec$. Since the LSPM-north is
statistically complete in high proper motion stars down to an apparent
magnitude $V=19$, our census of wide binaries is complete to a
distance of 100 pc for companions with absolute magnitude
$M_V<14$. This high completeness yields the most thorough search for
wide companions to {\em Hipparcos} stars conducted to date.

We find that the observed luminosity function of the faint secondaries
appears to be biased against low-luminosity stars, when compared with
the luminosity function of the field, single star population. Most of
the {\em Hipparcos} primaries are F-G dwarfs with absolute magnitudes
around $M_V=5$. The secondaries, on the other hand, are dominated by
fainter K-M dwarfs, with a range of absolute magnitude peaking at
$M_V=12$. While fainter, lower-mass companions clearly dominate over
companions of similar masses and luminosities, one finds only half the
number of faint secondaries as one would expect if both primaries and
secondaries were drawn from the field luminosity function. The bias
against low-mass secondaries also appears to get more severe at lower
secondary masses.

We further observe a distribution in the projected separations $s$
which is consistent with {\em \"Opik's law} ($f(s) ds \sim s^{-1}$ ds)
for systems with orbital separations up to $\approx3,500$ AU. Beyond that
range, the frequency of wide binaries decreases according to a power
law $f(s) ds \sim s^{-1.6}$ ds. Our data thus provides clear evidence
for a break in {\em \"Opik's law} at large angular separations, and we
suggest that the break occurs as interactions between neighboring
systems either prevents the formation of very wide binary systems or
tends to disrupt them after they are formed.

Measurements of the distribution of orbital separations remain
relatively crude. A more detailed analysis of the binary frequency,
which holds the promise to provide stringent constraints on multiple
star formation models, will require much larger samples of CPM
pairs. At this time, we plan to first expand our census to the
southern hemisphere using the upcoming LSPM-south catalog (L\'epine \&
Shara, in preparation) which will be the analog of the LSPM-north for
the southern declinations. This will double our census CPM pairs. The
census of wide binaries should also be expanded further by extending
the search to smaller proper motion regimes. Such a survey of pairs of
stars with low proper motions has recently been demonstrated by
\citet{G04}, who identified 705 wide binaries among stars with proper
motions ($\mu<0.2\arcsec$ yr$^{-1}$) based on an analysis of the UCAC2
astrometric catalog. The {\it Hipparcos} catalog has 16,753 objects
within 100 parsecs and parallaxes better than $10\%$. However, only
6,202 of these have proper motions $\mu>0.15\arcsec yr^{-1}$, the
current range of our proper motion survey. However, 13,368 of these
nearby {\it Hipparcos} stars have proper motion $\mu>0.05\arcsec
yr^{-1}$, so a catalog of stars with $\mu>0.05\arcsec
yr^{-1}$ would potentially double the census of wide binaries. The
census could further be expanded by investigating {\it Hipparcos}
stars at larger recorded distances, adding up to 10,000 targets in the
100-200pc range.

It would also be desirable to break the low angular separation limit
$\Delta \theta>20\arcsec$ of the photographic plate-based surveys. A
look at Fig.7 is highly suggestive of the existence of significant
numbers of faint companions with small angular separations. With CCD
imaging and adaptive optics systems, it should be possible to perform
a complete census of CPM companions of {\it Hipparcos} stars down to
angular separations of a fraction of an arcsecond. This would be
particularly useful in extending the census to pairs with smaller
orbital separations (s$<1,000$ AU) and verify whether this regime can
be consistently modeled with {\it \"Opik's Law}.

\acknowledgments

{\bf Acknowledgments}

SL gratefully acknowledges support from Hilary Lipsitz, and from the
American Museum of Natural History.


\clearpage
\begin{landscape}
\begin{deluxetable}{lrrrrrrrrrclrrrrrrrrrr}
\tabletypesize{\scriptsize}
\tablewidth{0pc}
\tablecolumns{22}
\tablecaption{Common proper motion systems with 2 or more {\it Hipparcos} stars\tablenotemark{a}}
\tablehead{
\multicolumn{11}{c}{Primary} & \multicolumn{10}{c}{Secondary/Tertiary} \\
\cline{1-10} \cline{12-21} \\
LSPM & HIP & R.A. & Decl. & $\mu_{R.A.}$ & $\mu_{Decl.}$ & V  & V-J & $\pi$ & $\pi_{err}$ & $\Delta\theta$ & 
LSPM & HIP & R.A. & Decl. & $\mu_{R.A.}$ & $\mu_{Decl.}$ & V  & V-J & $\pi$ & $\pi_{err}$ & notes\tablenotemark{b}\\
  &  &   &   & $\arcsec$ yr$^{-1}$ & $\arcsec$ yr$^{-1}$ & mag & mag & $mas$ & $mas$      & $\arcsec$ &
  &  &   &   & $\arcsec$ yr$^{-1}$ & $\arcsec$ yr$^{-1}$ & mag & mag & $mas$ & $mas$      & \\
 (1) &  (2) &  (3) &  (4) &  (5) &  (6) &  (7) &  (8) &  (9) & (10) &
(11) & (12) & (13) & (14) & (15) & (16) & (17) & (18) & (19) & (20) & (21) & (22)
}
\startdata
J0005+1804W &    495 &    1.481144&  18.075895&  -0.146&  -0.146&  9.22&  2.13&   25.8&  2.1&    3.5& J0005+1804E &    495 &    1.481890&  18.075220&  -0.154&  -0.154&  9.55&  2.43&   25.8&  2.1&  T \\
J0005+1814  &    493 &    1.478123&  18.235014&  -0.149&  -0.151&  7.47&  1.14&   26.2&  0.8&  572.9& J0005+1804W &    495 &    1.481144&  18.075895&  -0.146&  -0.146&  9.22&  2.13&   25.8&  2.1&  T \\
J0005+4548N &    473 &    1.420888&  45.812080&   0.879&  -0.154&  8.83&  2.73&   85.1&  2.7&  328.0& J0005+4547  &    428 &    1.295397&  45.786568&   0.870&  -0.151& 10.05&  3.35&   87.0&  1.4&  T \\
J0005+4548N &    473 &    1.420888&  45.812080&   0.879&  -0.154&  8.83&  2.73&   85.1&  2.7&    6.1& J0005+4548S &    473 &    1.420823&  45.810383&   0.839&  -0.162&  8.97&  2.83&   85.1&  2.7&  T \\
J0032+6714N &   2552 &    8.122603&  67.235619&   1.741&  -0.247& 10.49&  3.65&   98.7&  3.4&    3.7& J0032+6714S &   2552 &    8.123230&  67.234619&   1.705&  -0.315& 12.19&  5.02&   98.7&  3.4&  \\
J0036+2959S &   2844 &    9.009863&  29.993027&   0.191&  -0.403&  8.32&  1.29&   18.9&  1.7&    6.1& J0036+2959N &   2844 &    9.010551&  29.994621&   0.177&  -0.388&  8.98&  1.19&   18.9&  1.7&   \\
J0049+5748  &   3821 &   12.276213&  57.815186&   1.087&  -0.560&  3.45&  1.34&  168.0&  0.6&   12.8& J0049+5749  &   3821 &   12.271529&  57.817715&   1.105&  -0.493&  7.36&  0.19&  168.0&  0.6&   \\
J0105+1523N &   5110 &   16.374641&  15.390063&   0.009&  -0.196&  9.12&  2.00&   37.6&  2.0&    6.1& J0105+1523S &   5110 &   16.373962&  15.388507&   0.006&  -0.194&  9.80&  2.65&   37.6&  2.0&  GC \\
J0120+3637N &   6240 &   20.018244&  36.630920&  -0.103&  -0.142&  8.87&  0.96&   15.4&  2.0&    4.9& J0120+3637S &   6240 &   20.018011&  36.629581&  -0.109&  -0.154&  9.73&  1.83&   15.4&  2.0&   \\
J0122+1245S &   6431 &   20.652483&  12.750941&   0.400&   0.008&  9.51&  1.65&   23.8&  2.1&    5.8& J0122+1245N &   6431 &   20.653130&  12.752419&   0.403&   0.009& 11.75&  3.07&   23.8&  2.1&   \\
J0125+3132E &   6675 &   21.410614&  31.546171&   0.155&  -0.050&  7.99&  1.15&   17.3&  1.0&   56.7& J0125+3133  &   6668 &   21.392822&  31.550421&   0.158&  -0.048&  8.03&  1.10&   16.6&  1.5&  T \\
J0125+3133  &   6668 &   21.392822&  31.550421&   0.158&  -0.048&  8.03&  1.10&   16.6&  1.5&    4.6& J0125+3132W &   6668 &   21.391500&  31.549845&   0.153&  -0.048& 11.79&  3.71&   16.6&  1.5&  T
\enddata
\tablenotetext{a}{The complete table of 168 lines is available in the 
electronic version of the Astronomical Journal. The first 12 lines are
shown here as a guide to the contents of the full table.}
\tablenotetext{b}{T: star in a common proper motion triple system, Q:
star in a common proper motion quadruple system, GC: star listed in
the \citet{GC04} catalog of common proper motion double.}
\end{deluxetable}
\clearpage
\end{landscape}

\clearpage
\begin{landscape}
\begin{deluxetable}{lrrrrrrrrrclrrrrrrl}
\tabletypesize{\scriptsize}
\tablewidth{0pc}
\tablecolumns{19}
\tablecaption{Common proper motion systems with only one {\it Hipparcos} stars, uncertain companionship}
\tablehead{
\multicolumn{10}{c}{Primary} & \multicolumn{8}{c}{Secondary/Tertiary} \\
\cline{1-10} \cline{12-18} \\
LSPM & HIP & R.A. & Decl. & $\mu_{R.A.}$ & $\mu_{Decl.}$ & V  & V-J & $\pi$ & $\pi_{err}$ & $\Delta\theta$ &
LSPM & R.A. & Decl. & $\mu_{R.A.}$ & $\mu_{Decl.}$ & V  & V-J & notes\\
  &  &   &   & $\arcsec$ yr$^{-1}$ & $\arcsec$ yr$^{-1}$ & mag & mag & $mas$ & $mas$ & &
  &  &   & $\arcsec$ yr$^{-1}$ & $\arcsec$ yr$^{-1}$ & mag & mag & \\
 (1) &  (2) &  (3) &  (4) &  (5) &  (6) &  (7) &  (8) &  (9) & (10) &
(11) & (12) & (13) & (14) & (15) & (16) & (17) & (18) & (19) 
 }
\startdata
J0005+1804W &    495&     1.481144&  18.075895&  -0.146&  -0.146&  9.22&  2.13&   25.8&  2.1& 1480.7& J0004+1803 &     1.049166&  18.052927&  -0.146&  -0.146& 16.58&  2.51&  background/subdwarf \\
J0300+0517  &  14054&    45.240772&   5.287946&   0.190&  -0.094&  8.14&  1.58&   11.7&  1.3&  359.4& J0300+0520 &    45.152958&   5.336117&   0.189&  -0.090& 19.10&  3.99&  background/subdwarf \\
J0316+5810N &  15220&    49.058128&  58.168789&   0.436&  -0.305& 11.25&  3.75&   74.3&  6.3& 1168.9& J0315+5751 &    48.872692&  57.859173&   0.467&  -0.237& 15.57&  4.45&  background/subdwarf \\
J0320+4358N &  15591&    50.203442&  43.978798&   0.137&  -0.118&  8.82&  1.25&   15.1&  1.2&   44.8& J0320+4358S&    50.212368&  43.968155&   0.137&  -0.118& 19.70&  3.60&  background/subdwarf \\
J0450+6319N &  22498&    72.604561&  63.332939&   0.220&  -0.196&  9.78&  2.23&   42.6& 17.8&   40.8& J0450+6319S&    72.589020&  63.324009&   0.236&  -0.196& 12.70&  2.67&  background/subdwarf \\
J0722+0849S &  35756&   110.631073&   8.820297&   0.154&  -0.265& 10.46&  1.14&    5.9&  2.1&   34.0& J0722+0849N&   110.624702&   8.827321&   0.149&  -0.267& 16.66&  2.70&  subdwarf \\
J0734+3153E &  36850&   113.650436&  31.888498&  -0.200&  -0.146&  2.96&  1.63&   63.3&  1.2&   70.5& J0734+3152 &   113.656021&  31.869511&  -0.203&  -0.094&  9.63&  3.56&  foreground/double \\
J0734+3153W &  36850&   113.649521&  31.888388&  -0.174&  -0.102&  1.88& -1.93&   63.3&  1.2&   70.8& J0734+3152 &   113.656021&  31.869511&  -0.203&  -0.094&  9.63&  3.56&  foreground/double \\
J0821+3418W &  40942&   125.336388&  34.309975&  -0.119&  -0.111&  8.52&  0.55&   23.1&  1.2&   72.9& J0821+3418N&   125.360886&  34.310619&  -0.111&  -0.116& 17.46&  2.52&  background/subdwarf \\
J1106+1415  &  54299&   166.631561&  14.262381&   0.208&  -0.314&  9.32&  2.24&    9.3&  2.2&  669.3& J1105+1412E&   166.447083&  14.211407&   0.197&  -0.327& 18.80&  5.96&  foreground/double \\
J1425+5151  &  70497&   216.299164&  51.850765&  -0.237&  -0.399&  4.05&  0.87&   68.6&  0.6&   69.5& J1425+5149 &   216.298294&  51.831474&  -0.243&  -0.404& 13.23&  5.35&  foreground/double \\
J1507+7612  &  74045&   226.984421&  76.200737&  -0.131&   0.164&  8.73&  1.90&   34.0&  0.7&  116.5& J1507+7613 &   226.988403&  76.233093&  -0.123&   0.157& 14.27&  5.04&  foreground/double \\
J1543+1101S &  76972&   235.753250&  11.017673&  -0.160&   0.163&  9.02&  1.07&    7.9&  1.7&   43.3& J1543+1101N&   235.751511&  11.029566&  -0.144&   0.143& 17.98&  5.32&  GC,foreground/double \\
J1604+3909E &  78775&   241.236649&  39.156509&  -0.572&   0.052&  6.66&  1.48&   69.6&  0.6&   70.3& J1604+3909W&   241.211884&  39.160019&  -0.547&   0.055& 12.86&  2.96&  GC,background/subdwarf \\
J1605+3500  &  78859&   241.468567&  35.006042&  -0.059&  -0.251&  8.06&  1.15&   17.1&  0.8&   42.6& J1605+3459 &   241.456970&  34.999004&  -0.041&  -0.242& 15.32&  3.06&  background/subdwarf \\
J1746+2743E &  86974&   266.614655&  27.720701&  -0.310&  -0.750&  3.41&  1.54&  119.1&  0.6&   34.2& J1746+2743W&   266.604828&  27.716873&  -0.333&  -0.753& 10.10&  4.33&  GC,foreground/double \\
J1803+8408N &  88435&   270.834686&  84.147728&  -0.125&   0.137& 11.23&  2.08&   17.5&  3.3&   30.9& J1803+8408S&   270.860107&  84.139549&  -0.123&   0.135& 17.18&  3.39&  background/subdwarf \\
J2109+5421  & 104426&   317.306641&  54.353336&   0.184&  -0.071&  9.87&  1.92&   22.2&  1.1&  113.7& J2109+5422 &   317.330933&  54.381577&   0.191&  -0.073& 14.74&  4.73&  foreground/double \\
J2114+3802  & 104887&   318.697754&  38.045372&   0.159&   0.431&  3.74&  0.86&   47.8&  0.6&  534.3& J2115+3804 &   318.881775&  38.077732&   0.216&   0.453& 16.02&  3.05&  background/subdwarf \\
J2227+4029  & 110865&   336.924835&  40.484993&   0.170&   0.035& 10.82&  1.84&   13.1&  2.4&  116.7& J2227+4027 &   336.940277&  40.454769&   0.162&   0.027& 17.77&  5.58&  foreground/double \\
\enddata
\end{deluxetable}
\clearpage
\end{landscape}

\clearpage
\begin{landscape}
\begin{deluxetable}{lrrrrrrrrrclrrrrrrr}
\tabletypesize{\scriptsize}
\tablewidth{0pc}
\tablecolumns{19}
\tablecaption{Common proper motion systems with only one {\it Hipparcos} stars\tablenotemark{a}}
\tablehead{
\multicolumn{10}{c}{Primary} & \multicolumn{8}{c}{Secondary/Tertiary} \\
\cline{1-10} \cline{12-18} \\
LSPM & HIP & R.A. & Decl. & $\mu_{R.A.}$ & $\mu_{Decl.}$ & V  & V-J & $\pi$ & $\pi_{err}$ & $\Delta\theta$ &
LSPM & R.A. & Decl. & $\mu_{R.A.}$ & $\mu_{Decl.}$ & V  & V-J & notes\tablenotemark{b}\\
  &  &   &   & $\arcsec$ yr$^{-1}$ & $\arcsec$ yr$^{-1}$ & mag & mag & $mas$ & $mas$ & &
  &  &   & $\arcsec$ yr$^{-1}$ & $\arcsec$ yr$^{-1}$ & mag & mag & \\
 (1) &  (2) &  (3) &  (4) &  (5) &  (6) &  (7) &  (8) &  (9) & (10) &
(11) & (12) & (13) & (14) & (15) & (16) & (17) & (18) & (19) 
 }
\startdata
J0006+0549W &    538 &    1.632458 &  5.825399 &  0.195&   0.040& 11.05&  1.84&   13.3&  2.4&   23.4& J0006+0549E&     1.637383&   5.821126&   0.195&   0.041& 16.63&  4.73&  \\
J0009+2516E &    754 &    2.315609 & 25.281969 &  0.173&  -0.149&  7.78&  1.29&   22.1&  2.3&   29.6& J0009+2516W&     2.307973&  25.277500&   0.166&  -0.145& 10.94&  2.10&  GC\\
J0009+2739  &    731 &    2.262073 & 27.651577 &  0.215&   0.144& 11.80&  2.38&   22.5&  3.0&   68.9& J0008+2739 &     2.248174&  27.666235&   0.215&   0.144& 13.70&  3.27&  GC\\
J0010+4806W &    889 &    2.733613 & 48.110405 &  0.169&   0.000&  8.41&  1.04&   16.5&  0.9&   23.9& J0010+4806E&     2.740871&  48.114941&   0.168&   0.003& 10.72&  1.87&  \\
J0012+2142N &   1006 &    3.139525 & 21.713451 &  0.183&  -0.289& 11.91&  3.07&   34.5& 11.7&   29.0& J0012+2142S&     3.143580&  21.706341&   0.187&  -0.294& 13.65&  3.99&  \\
J0013+8039W &   1092 &    3.411198 & 80.665802 &  0.253&   0.185& 11.15&  3.39&   51.1&  3.3&   12.9& J0013+8039E&     3.429084&  80.663689&   0.259&   0.193& 16.88&  5.94&  GC\\
J0015+5304S &   1224 &    3.811452 & 53.074158 &  0.220&   0.044& 10.41&  1.57&   12.2&  2.1&   18.8& J0015+5304N&     3.810659&  53.079353&   0.232&   0.005& 14.34&  3.52&  GC\\
J0016+1951W &   1295 &    4.060917 & 19.860455 &  0.709&  -0.748& 11.85&  3.97&   49.9& 21.6&   25.1& J0016+1951E&     4.067259&  19.864059&   0.709&  -0.748& 13.55&  4.66&  \\
J0018+4401W &   1475 &    4.595309 & 44.022949 &  2.888&   0.409&  8.15&  2.90&  280.3&  1.0&   36.0& J0018+4401E&     4.607785&  44.027344&   2.912&   0.351& 11.12&  4.33&  GC\\
J0023+7711W &   1860 &    5.870120 & 77.189301 & -0.838&   0.045& 11.53&  3.49&   50.7&  2.7&   11.3& J0023+7711E&     5.882650&  77.190758&  -0.838&   0.045& 15.17&  5.24&  GC\\
J0033+4443  &   2600 &    8.255900 & 44.730080 &  0.222&  -0.044& 10.29&  1.48&    9.5&  1.6&   28.9& J0033+4444 &     8.261676&  44.736973&   0.220&  -0.049& 16.85&  0.18&  GC\\
J0038+4259W &   3022 &    9.621646 & 42.999924 &  0.192&  -0.083&  9.95&  1.54&   19.9&  1.9&   53.1& J0038+4259E&     9.638277&  42.991577&   0.194&  -0.079& 13.54&  3.05&  GC\\
\enddata
\tablenotetext{a}{The complete table of 364 lines is available in the 
electronic version of the Astronomical Journal. The first 12 lines are
shown here as a guide to the contents of the full table.}
\tablenotetext{b}{T: star in a common proper motion triple system, Q:
star in a common proper motion quadruple system, GC: star listed in
the \citet{GC04} catalog of common proper motion double.}
\end{deluxetable}
\clearpage
\end{landscape}

\end{document}